\documentclass[]{article}
\usepackage{graphicx}
\usepackage{xcolor}
\usepackage{amsfonts}
\usepackage{amssymb}
\usepackage{pxfonts}
\def\ligne#1{\hbox to\hsize{#1}}
\def\leurre{\noindent\leftskip0pt\small\baselineskip 10pt}
\newtheorem{thm}{\textrm{\sc Theorem}}
\newtheorem{cor}{\textrm{\sc Corollary}}
\newtheorem{lemm}{\textrm{\sc Lemma}}
\newtheorem{fig}{\textrm{Figure}}
\newtheorem{tab}{\textrm{Table}}
\newtheorem{algo}{\textrm{Algorithm}}

\newcounter{laform}
\author{Maurice {\sc Margenstern}}
\title{About Fibonacci trees. 
\begin{center} $-$ II $-$ :\\ generalized Fibonacci trees\end{center}}

\begin{document}
\maketitle
\begin{abstract}
In this second paper, we look at the following question: are the properties of the 
trees associated to the tilings $\{p,4\}$ of the hyperbolic plane still true if 
we consider a finitely generated tree by the same rules but
rooted at a black node? The direct answer is no, but new properties arise, no more 
complex than in the case of a tree rooted at a white node, and worth of interest.
The present paper is an extension of the previous paper~\cite{mmblacktree}.
\end{abstract}

\section{Introduction}
\def\zz#1{{\footnotesize\tt #1}}

   The existence of a tree generating the pentagrid, {\it i.e.} the tiling $\{5,4\}$
of the hyperbolic plane generalizes to the tilings $\{p,4\}$ in the same plane.
This paper can be seen as an extension of the previous paper~\cite{mmblacktree}
which investigated the following question: are the properties of the Fibonacci tree of
the pentagrid still true if we consider a tree rooted at a black node?
As in~\cite{mmblacktree}, we shall see that preferred son property is no more true
but that new properties, sightly more complex ones, arise in their place.

   In this new setting, we generalize what we called the golden sequence 
in~\cite{mmblacktree} to what could be called a generalized Fibonacci sequence
but which we shall call the {\bf metallic sequences} which we define in
section~\ref{trees}: the Fibonacci sequence is connected with the golden number
which is the root of a polynomial whose form is a particular case of the polynomials
we shall meet in the paper. In that section too, we remind the reader some properties 
about
infinite trees and numbers connected with the rules which defines those trees.
We shall define two kinds of them. One kind is studied in Section~\ref{white_metal},
the other in Section~\ref{black_metal} where we investigate the properties of
a black metallic tree. In Subsections~\ref{stilings} and~\ref{bpentahepta}  
we indicate the connection of those
trees with two infinite families of tilings of the hyperbolic plane.
In Section~\ref{scompnumwb}, we compare the properties studied in 
Sections~\ref{white_metal} and~\ref{black_metal}, giving an explanation to 
the differences we observed. Section~\ref{conclude} concludes the paper.

\section{The metallic trees}\label{trees}

   In Subsection~\ref{smetalsprop}, we remind the reader with general definitions
about infinite trees with finite branching. Then, in Subsection~\ref{smetalnum}
we introduce the definition of the metallic trees and then of the
metallic sequences we associate to them. In that section too, in 
Subsection~\ref{smetalrepr}
we look at the metallic representation of the positive numbers
as sums of terms of the metallic sequence
and the connections of those numbers
with the trees. 
We define addition and subtraction
on the representations of metallic numbers in Subsection~\ref{sopermcodes}, which will
help us to establish the properties investigated in Sections~\ref{white_metal}
and~\ref{black_metal}.

\subsection{Preliminary properties}\label{smetalsprop}

Consider an infinite tree~$\cal T$ with finite branching at each node. Number the nodes
from the root which receives~1, then, level by level and, on each level, from left to 
right with the conditions that for each node, the numbers of its sons are consecutive
numbers. We then say that $\cal T$ is {\bf numbered} or that it is endowed with
its {\bf natural numbering}. In what follows, we shall consider
numbered trees only. Clearly, a sub-tree $\cal S$ of $\cal T$ can also be numbered
in the just above described way but it can also be numbered by the numbers of its
nodes in $\cal T$. In that case, a node~$\nu$ may receive two numbers: $n_{\cal S}$,
the number defined in~$\cal S$ as a numbered tree and $n_{\cal T}$, its number as a
node of~$\cal T$. A node may have no son, it is then called a {\bf leaf}.
A {\bf path} from~$\mu$ to~$\nu$ is a finite sequence of nodes 
\hbox{$\{\lambda_i\}_{i\in[0..k]}$}, if it exists, such that $\lambda_0=\mu$, 
$\lambda_k=\nu$ and, for all $i$ with \hbox{$i\in[0..k$$-$$1]$}, $\lambda_{i+1}$ is a 
son of~$\lambda_i$.  
A {\bf branch} of $\cal T$ is a maximal finite or infinite sequence of paths~$\{\pi_i\}$ 
from the root of~$\cal T$ to nodes of that tree such that for all~$i$, $j$, 
$\pi_i\subseteq\pi_j$ or $\pi_j\subseteq\pi_i$. Accordingly, a branch connects the root
to a leaf or it is infinite. It is clear that for any node, they are connected to 
the root by a unique path. The {\bf length} of the path from a node to one of its son 
is always~1. If the length of a path from~$\mu$ to~$\nu$ is~$k$, the length of the
path from~$\mu$ to any son of~$\nu$, assuming that $\nu$ is not a leaf, is $k$+1.
The length of the path leading from the root to a node~$\nu$ of~$\cal T$ is
called the {\bf distance} of~$\nu$ to the root~$\rho$ and it is denoted by
dist$\{\rho,\nu\}$. We also define \hbox{dist$\{\rho,\rho\}=0$}. 
The {\bf level}~$k$ of~$\cal T$ is the set of its nodes which are at the distance~$k$
from its root. Denote it by \hbox{${\cal L}_{k,\cal T}$}. 
Define~${\cal T}_n$ as the set of levels~$k$ of~$\cal T$ with
\hbox{$k\leq n$}. Say that the {\bf height} of ${\cal T}_n$ is~$n$.
By definition, ${\cal T}_n$ is a sub-tree of~$\cal T$. 
For each node~$\nu$ of~$\cal T$, $\lambda_{\cal T}(\nu)$ is its level 
in~$\cal T$, {\it i.e.} its distance from the root, and $\sigma_{\cal T}(\nu)$ is the 
number of its sons. Clearly, if $\nu \in {\cal T}_n$ and if $\lambda_{\cal T}(\nu)=n$, 
then $\sigma(\nu)=0$. If $\cal S$ is a sub-tree of $\cal T$, denote it
by ${\cal S}\lhd\cal T$, and if $\nu\in\cal S$,
then $\lambda_{\cal S}(\nu)\leq\lambda_{\cal T}(\nu)$ and the numbers may be not equal.

Consider two infinite numbered trees $T_1$ and $T_2$. 
Say that $T_1$ and $T_2$ are {\bf isomorphic} if 
there is a bijection~$\beta$ from~$T_1$ onto~$T_2$ such that:
\vskip 5pt
\ligne{\hfill
$\vcenter{\vtop{\leftskip 0pt\hsize=150pt
\ligne{$f(n_{T_1})=n_{T_2}$ for any $n\in\mathbb N$.\hfill}
\ligne{$\lambda_{T_2}(f(n_{T_1}))=\lambda_{T_1}(n)$.\hfill}
\ligne{$\sigma_{T_2}(f(n_{T_1}))=\sigma_{T_1}(n)$.\hfill}
}}$
\hfill$(0)$\hskip 10pt}
\vskip 5pt
% définir \cal W_n, \cal B_n, \cal C_n, isomorphisme, m_n, M_n, b_n...

\subsection{Metallic trees and metallic numbers}\label{smetalnum}

    We call {\bf metallic tree} a finitely generated tree with two kinds of nodes,
{\bf black} nodes and {\bf white} ones whose generating rules are:
\vskip 5pt
\ligne{\hfill$B\rightarrow BW^{p-4}$ and $W\rightarrow BW^{p-3}$.\hfill (1)\hskip 10pt}
\vskip 5pt
\noindent
with \hbox{$p\geq5$}.

The property for a node to be white or black is called its {\bf status}.

We shall mainly investigate two kinds of infinite metallic trees.
When the root of the tree is a white, black node, we call such a metallic tree a 
{\bf white}, {\bf black metallic tree} respectively. We denote the
infinite white metallic tree by $\cal W$ and we endow it with its natural numbering.
We do the same with the infinite black metallic tree $\cal B$. Note
that we can construct a bijective morphism between $\cal B$ and a part $\mathbb B$ 
of~$\cal W$ as follows. The morphism is the identity on~$\mathbb B$
and we fix the following conditions: 
\vskip 5pt
\ligne{\hfill
\vtop{\leftskip 0pt\hsize=200pt
\ligne{\hfill$\sigma_{\mathbb B}(1) = \sigma_{\cal W}(1)-1$,\hfill}
\ligne{\hfill$\sigma_{\mathbb B}(n) = \sigma_{\cal W}(n)$, for all positive integer~$n$.
\hfill}
}
\hfill}
\vskip 5pt
\noindent
Moreover, the nodes numbered by~$n\in [1..p$$-$$2]$ in $\cal W$ also belong
to $\mathbb B$ and receive the same numbers in the natural numbering of $\mathbb B$.
This morphism allows us to identify $\mathbb B$ with $\cal B$, so that in our
sequel, we shall speak of $\cal B$ only. From what we just said, it is plain that
for a node $\nu\in \cal B$, if $\nu_{\cal B} > p$$-$2, then
$\nu_{\cal B} < \nu_{\cal W}$. We shall look closer to the connection between
$\nu_{\cal B}$ and $\nu_{\cal W}$ in Section~\ref{scompnumwb}.

Before turning to the properties of $\cal W$ and $\cal B$ separately, we shall
study the connection of the numbering with respect to properties which are
associated with the rules~$(1)$.

To that purpose let $m_n$, $b_n$ be the number of nodes on ${\cal L}_{n,\cal W}$ 
and ${\cal L}_{n,\cal B}$ respectively. We also define $M_n$, and $B_n$ as the number
of nodes of ${\cal W}_n$ and ${\cal B}_n$ respectively.

   The connection with the metallic sequence first appear when we count the number of 
nodes which lay at the same level of the tree. For a white metallic tree, we have the
following property:

\begin{thm}{\rm\cite{mmJUCStools}}\label{tmetallevelw}
Consider the numbers $m_n$ defined as the number of nodes on ${\cal L}_{n,\cal W}$,
where $\cal W$ is the white metallic tree. The numbers $m_n$ satisfy the
following induction equation:
\vskip 5pt
\ligne{\hfill$m_{n+2} = (p$$-$$2)m_{n+1}-m_n$ with $m_0=1$ and $m_{-1}=0$.\hfill
$(2)$\hskip 10pt}
\vskip 5pt
\noindent
We call {\bf white metallic sequence} the sequence $\{m_n\}_{n\in\mathbb N}$.
\end{thm}

\noindent
Proof. Note that each node gives rise to $p$$-$2 sons if it is white, to $p$$-$3 of them
if it is black. Now, each node gives rise to exactly one black son. Accordingly,
if $m_n$ is the number of nodes on the level~$n$, we have that 
\hbox{$m_{n+2} = (p$$-$2$)m_{n+1}-m_n$} as the number of black nodes on the level~$n$+1
is $m_n$ from what was just said and as in considering \hbox{$(p$$-$2$)m_{n+1}$},
we count twice the black nodes yielded by the black nodes of the level~$n$+1.
\hfill$\Box$

As the black metallic tree is defined by the same rules, we may conclude that
the same equation rules the sequence $\{b_n\}_{n\in\mathbb N}$:

\begin{thm}\label{tmetallevelb}
The sequence $\{b_n\}_{n\in\mathbb N}$ of the number of nodes on ${\cal L}_{n,\cal B}$
satisfies the equation:
\vskip 5pt
\ligne{\hfill$b_{n+2} = (p$$-$$2)b_{n+1}-b_n$ with $b_1=p$$-$$3$ and $b_0=1$.\hfill
$(3)$\hskip 10pt}
\vskip 5pt
We call {\bf black metallic sequence} the sequence $\{b_n\}_{n\in\mathbb N}$. 
\end{thm}

Note that we could define the white metallic sequence by the initial conditions
$m_1=p$$-$2 and $m_0=1$. In our sequel we shall say {\bf metallic sequence} instead
of {\bf white metallic sequence} for a reason which will be made more clear in a while.

Before turning to the properties of the integers with respect to the metallic numbers,
we have to consider the numbers $M_n$ and $B_n$ already introduced with respect 
to the finite trees ${\cal W}_n$ and ${\cal B}_n$.

\begin{thm}{\rm\cite{mmbook1}}\label{theadlevelwb}
On the level~$k$ of $\cal W$, with non-negative~$k$, the rightmost node has the number
$M_k$, so that the leftmost node on the level~$k$$+$$1$
has the number $M_k$$+$$1$.

On the level~$k$ of $\cal B$ with non-negative~$k$, the rightmost node has the
number~$m_k$, so that the leftmost node on the level~$k$$+$$1$ has the
number~$m_k$$+$$1$.

The sequence $\{M_n\}_{n\in\mathbb N}$ satisfies the following induction equation:
\vskip 5pt
\ligne{\hfill$M_{n+2}=(p$$-$$2$$)M_{n+1}-M_n+1$,\hfill 
$(4)$\hskip 10pt}
\vskip 5pt
\noindent
with the initial conditions $M_0=1$ and $M_{-1}=0$, while the sequence 
$\{B_n\}_{n\in\mathbb N}$ satisfy the equation~$(2)$ with the same
initial conditions, which means that \hbox{$B_n=m_n$} for any non-negative~$n$.
We also have:
\vskip 5pt
\ligne{\hfill $M_{n+1} = B_{n+1}+M_n$\hskip 10pt and
\hskip 10pt $m_{n+1} = b_{n+1}+m_n$\hfill $(5)$\hskip 10pt}
\end{thm}

\noindent
Proof. Consider the numbers $M_n$. We can write:
\vskip 5pt
\ligne{\hskip 20pt $M_{n+2}=\displaystyle{\sum\limits_{i=0}^{n+2} m_i}
=\displaystyle{\sum\limits_{i=0}^n m_{i+2}}+m_1+m_0
=\displaystyle{\sum\limits_{i=0}^n m_{i+2}}+p$$-$1
\hfill}
\ligne{\hskip 20pt $=(p$$-$$2)\displaystyle{\sum\limits_{i=0}^n m_{i+1}}
-\displaystyle{\sum\limits_{i=0}^n m_i}+p$$-$1
$=(p$$-$$2)(M_{n+1}$$-$$M_0)-M_n+p$$-$1
\hfill}
\noindent
which is the equation $(4)$.

In the case of the black metallic sequence, the same computation shows
that \hbox{$b_1$+$b_0=p$$-$2} which is cancelled by the sum
$(p$$-$$2)\displaystyle{\sum\limits_{i=0}^n b_{i+1}}=(p$$-$$2)(B_{n+1}-B_0)$,
so that the sequence satisfies $(2)$ with the same initial conditions.
Another way to see that is to observe that
from the decomposition of \hbox{${\cal W}={\cal B}\cup \cal C$} with
\hbox{${\cal B}\cap{\cal C}=\emptyset$} we can see that
\hbox{${\cal W}_{n+1}={\cal B}_{n+1}\cup{\cal C}_n$}. Indeed,
${\cal C}$ is isomorphic to $\cal W$ if we take into account that the image of the root
of~$\cal C$ is the rightmost son of ${\cal L}_{1,\cal W}$, so that
\hbox{$M_{n+1}=B_{n+1}+M_n$}. Taking the trace of the decomposition 
\hbox{${\cal W}_{n+1}={\cal B}_{n+1}\cup{\cal C}_n$} on ${\cal L}_{n+1,\cal W}$,
we get that \hbox{$m_{n+1} = b_{n+1} + m_n$}.
Accordingly, $B_n$ is the difference of two terms of the sequence defined by~$(4)$,
so that the equation satisfied by~$B_n$ is obtained from~$(4)$ by cancelling the
term +1. So that $B_n$ satisfies (2) with the same conditions and so,
\hbox{$B_n=m_n$} for any~$n$ in~$\mathbb N$.\hfill $\Box$

\subsection{Metallic representation of the natural numbers and metallic codes
for the nodes of the metallic trees}~\label{smetalrepr}.

Let us go back to the sequence $\{m_n\}_{n\in\mathbb N}$ of metallic numbers.
It is clear that the sequence defined by~$(2)$ is increasing starting from~$m_1$:
from~$(2)$, we get that \hbox{$m_{n+2}> (p$$-$3$)m_{n+1}$} if we assume that
\hbox{$m_n<m_{n+1}$}. As \hbox{$p\geq5$}, we get that the sequence is increasing starting
from~$m_1$. Now, as the sequence is increasing, it
is known that any positive integer~$n$ can be written as a sum of distinct metallic 
numbers whose terms are defined by Theorem~\ref{tmetallevelw}:
\vskip 5pt
\ligne{\hfill$n=\displaystyle{\sum\limits_{i=0}^k a_im_i}$ with $a_i\in\{0..p$$-$$3\}$.
\hfill
(6)\hskip 10pt}
\vskip 5pt
\noindent
The sum of $a_im_i$'s in~$(6)$ is called the {\bf metallic representation} of~$n$
and the $m_i$'s in~$(6)$ are the {\bf metallic components} of~$n$.

\def\bbzz{\hbox{\bf b0}}
\def\bbuu{\hbox{\bf b1}}
\def\wwzz{\hbox{\bf w0}}
\def\wwaa{\hbox{\bf wa}}
\def\bbb{\hbox{\bf b}}
\def\aa{\hbox{\bf a}}
\def\zzz{\hbox{\bf 0}}
\def\uu{\hbox{\bf 1}}
\def\ddd{\hbox{\bf d}}
\def\ccc{\hbox{\bf c}}
\def\eee{\hbox{\bf e}}
\def\www{\hbox{\bf w}}
\def\ppp{\hbox{\bf p}}
\def\numq{\hbox{\bf 4}}
\def\numd{\hbox{\bf 2}}
\def\numt{\hbox{\bf 3}}
\def\sympl{\hbox{\bf +}}
From now on, we use bold characters for the digits of a metallic representation of
a number. In particular, we define \ddd{} to represent $p$$-$3, 
\ccc{} to represent $p$$-$4 and \eee{} to represent $p$$-$5 when \hbox{$p>5$}. Of 
course, \zzz, \uu, \numd{} and \numt{} represent 0, 1, 2 and 3 respectively.

   First, note that the representation~$(6)$ is not unique.

\begin{lemm}{\rm \cite{mmgsJUCS,mmbook1}}\label{ltechforbid}
For any integers $n$ and $h$ with \hbox{$0\leq h\leq n$}, we have:
\vskip 5pt
\ligne{\hskip 20pt $(p$$-$$3)m_{n+1}+
	\displaystyle{\sum\limits_{k=h+1}^n\hbox{$(p$$-$$4)m_k$}}+(p$$-$$3)m_h$\hfill}
\ligne{\hskip 40pt
$=(p$$-$$3)m_{n+1}+
	\displaystyle{\sum\limits_{k=h+2}^n\hbox{$(p$$-$$4)m_k$}}+(p$$-$$3)m_{h+1}
-m_h+m_{h-1}$\hfill $(7)$\hskip 10pt}
\end{lemm}

\begin{cor}{\rm \cite{mmgsJUCS,mmbook1}}\label{ctechforbid}
For any positive integer~$n$, we have:
\vskip 5pt
\ligne{\hfill $(p$$-$$3)m_{n+1}+
	\displaystyle{\sum\limits_{k=1}^n\hbox{$(p$$-$$4)m_k$}}+(p$$-$$3)m_0
=m_{n+2}$\hfill $(8)$\hskip 10pt}
\end{cor}

\noindent
Proof.
By induction on the starting index in the summing sign, Lemma~\ref{ltechforbid} shows 
us that:
\vskip 5pt
\ligne{\hfill$(p$$-$$3)m_{n+1}+
\displaystyle{\sum\limits_{k=h+1}^n\hbox{$(p$$-$$4)m_k$}}+(p$$-$$3)m_h
=m_{n+2}-m_{h-1}$.\hfill$(9)$\hskip 10pt}
\vskip 5pt
\noindent
The corollary follows immediately from $(9)$ by making $h=0$
as, by assumption, $m_{-1}=0$.

\noindent
Proof of Lemma~\ref{ltechforbid}. 
We can see that the last term \hbox{$(p$$-$$3)m_h$} of the left-hand side of~$(7)$
can be developed as follows:
\vskip 5pt
\ligne{\hfill
\hbox{$(p$$-$$3)m_h = (p$$-$$2)m_h-m_h=m_{h+1}-m_h+m_{h-1}$}.\hfill}
\vskip 5pt
\noindent
Putting the right-hand side of that computation into the left-hand side member of~$(7)$
we get its right-hand side member.\hfill $\Box$

Let us write the $a_i$'s of~(5) as a word \hbox{\bf a$_k$..a$_1$a$_0$} which we call
a {\bf metallic word} for~$n$ as the digits $a_i$ which occur in~(5) 
are not necessarily unique for a given~$n$. 
They can be made unique by adding the following condition on the corresponding metallic
word for~$n$: the pattern \hbox{\bf dc$^\ast$d} is ruled out from
that word. It is called the {\bf forbidden pattern}. 
Lemma~\ref{ltechforbid} proves that property which is also proved 
in~\cite{mmgsJUCS,mmbook1}. We reproduced it here for the reader's convenience.

\def\sgn#1{\hbox{\bf sg}}
When a metallic representation for~$n$ does not contain the forbidden
pattern it is called the {\bf metallic code} of~$n$ which we denote by $[n]$. We shall 
write \hbox{$\nu = ([\nu])$} when we wish to restore the number from its metallic code. 
Let us call {\bf signature} of $\nu$ the rightmost digit of 
\hbox{\bf $[\nu]$ $=$ a$_k$$..$a$_1$a$_0$} and denote
it by \sgn{$(\nu)$}.
Let $\sigma_1$, $\sigma_2$, ..., $\sigma_k$ with $k=p$$-$2 or $k=p$$-$3 be the sons 
of~$\nu$. We call {\bf sons signature} of~$\nu$ the word \hbox{\bf s$_1$...s$_k$},
where \hbox{\bf s$_i$ $=$ \sgn{}$(\sigma_i)$}. 

\subsection{Operations on metallic codes}\label{sopermcodes}

We need to define additions and subtractions on metallic codes.
For the addition, we have the following algorithm:
\def\wwhile{{\bf while}}
\def\iff{{\bf if}}
\def\ffor{{\bf for}}
\def\endloop{{\bf end loop}}
\def\endif{{\bf end if}}
\def\iin{{\bf in}}
\def\faux{{\it false}}
\def\vrai{{\it true}}
\def\nnot{{\bf not}}
\def\lloop{{\bf loop}}
\def\tthen{{\bf then}}
\def\eelse{{\bf else}}
\def\sls{{\bf s}}
\def\cry{{\bf carry}}
\vskip 5pt
\vtop{
\begin{algo}\label{aaddmetal}
Adding two codes: {\bf a$_k$...a$_0$} and {\bf b$_k$...b$_0$}. The digits
are denoted by \aa$(i)$ and \bbb$(i)$, \aa{} and \bbb{} being seen as
tables. Also \sls{} is a table. We use global affectations to update \aa, \bbb,
\sls{} and the auxiliary table \cry{} in order to shorten the writing of the
algorithm.
\end{algo}
\vskip -8pt
\ligne{\hfill
\vtop{\leftskip 0pt\hsize=300pt
\hrule height 0.3 pt depth 0.3pt width \hsize
\vskip 5pt
\ligne{carry := 0; completed := \faux; test := 0; \hfill}
\ligne{\wwhile{} \nnot{} completed\hfill}
\ligne{\hskip 10pt \lloop{} \ffor{} i \iin{} [0..$k$]\hfill}
\ligne{\hskip 40pt \lloop{} \sls$(i)$ := \aa$(i)$ + \bbb$(i)$;\hfill}
\ligne{\hskip 70pt \iff{} \sls$(i)$ $\geq$ {\bf 10}\hfill}
\ligne{\hskip 90pt \tthen{} \cry($i$+1) := \cry($i$+1) + 1;\hfill}
\ligne{\hskip 110pt \iff{} $i > 0$\hfill}
\ligne{\hskip 130pt \tthen{} \cry($i$$-$1) := \cry($i$$-$1) + 1;\hfill}
\ligne{\hskip 110pt \endif{};\hfill}
\ligne{\hskip 100pt \sls$(i)$ := \sls$(i)$ $-$ {\bf 10};\hfill}
\ligne{\hskip 70pt \endif{};\hfill}
\ligne{\hskip 70pt test := $\displaystyle{\sum}$\cry$(i)$;\hfill}
\ligne{\hskip 70pt \iff{} test = 0\hfill}
\ligne{\hskip 90pt \tthen{} completed := \vrai;\hfill}
\ligne{\hskip 90pt \eelse{} \aa{} := \sls; \bbb{} := \cry; \cry{} := 0;\hfill}
\ligne{\hskip 70pt \endif{};\hfill}
\ligne{\hskip 40pt \endloop;\hfill}
\ligne{\hskip 10pt \endloop;\hfill}
\vskip 5pt
\hrule height 0.3 pt depth 0.3pt width \hsize
}
\hfill}
}
\vskip 5pt
\vskip 5pt
Of course, if the forbidden pattern occurs, we convert if to the correct form:
we replace the pattern by the same number of digits appending one to the first
digit which is on the left-hand side of the pattern. Note that several forbidden
patterns may occur in the result of Algorithm~\ref{aaddmetal}. Now,
that algorithm can be used to eliminate the occurrences of forbidden patterns 
in the metallic representation of a number.

To that purpose, note that the equation~$(2)$ can be rewritten as follows:
\vskip 5pt
\ligne{\hfill\bf  $($10$)($10$^{k+1}) = ($1010$^k)$ \hfill $(10)$\hskip 10pt}
\vskip 5pt
\noindent
as far as \hbox{\bf $($10$) = p$$-$2}, \hbox{\bf [$m_k$] $=$ 10$^k$} and, by
convention, \hbox{\bf $\nu$ $=$ $($[$\nu$]$)$}. Accordingly, appending
{\bf 10} at some place~$k$ involves a carry~1 on two places: $k$+1 and $k$$-$1,
except if the place is~0, in which case the carry applies to place~1 only.
It is the reason of the instructions managing the carry in Algorithm~\ref{aaddmetal}.
We already now from Corollary~\ref{ctechforbid}, that
\vskip 5pt
\ligne{\hfill\bf $($dc$^k$d$) = ($10$^{k+2})$\hfill}
\vskip 5pt
\noindent
In fact, the relation~$(9)$ given in the proof of Lemma~\ref{ltechforbid} can be
rewritten as:
\vskip 5pt
\ligne{\hfill\bf $($dc$^k$d0$^h) = ($10$^{k+2}$10$^{h-1})$. \hfill $(11)$\hskip 10pt}
\vskip 5pt
\noindent
which means that if the pattern \hbox{\bf dc$^k$d} occurs with its right-hand side~\ddd{}
at the place~$a$ and its left-hand side one at the place~$a$+$k$+1, increasingly 
numbering the places from right to left, the pattern is replaced by $k$+2 \zzz's
at the same place and a carry~\uu{} is put at the places $a$+$k$+2 and $h$$-$1.
As a consequence, if a forbidden pattern occurs among the metallic code of a number,
we replace the pattern by the needed number of~\zzz's at the same places and 
we add to that new number with the help of Algorithm~\ref{aaddmetal}, the number whose 
representation is given in the right hand-side part of~$(11)$. From this, we
can see that such an operation is repeated until no forbidden pattern occurs.

Now, we can turn to the subtraction of two numbers $a$ and $b$ performed on their 
metallic codes, which we may assume to be free of any forbidden pattern.

We decompose the subtraction of~$b$ from~$a$ into three parts. First, we check
whether \hbox{$a>b$} directly on {\bf [$a$]} and {\bf [$b$]}. If it is not
the case, the subtraction is not possible unless $a=b$ and the algorithm stops here. 
If it is the
case, by appending possible leading \zzz's, we may assume that {\bf [$a$]} and
{\bf [$b$]} have the same length. Let us consider the codes {\bf [$a$]} and
{\bf [$b$]} as tables. We denote by {\bf [$a$]$(i)$},
{\bf [$b$]$(i)$} the digit of {\bf [$a$]}, {\bf [$b$]} respectively, which occurs at the
place~$i$, where $i$ is the place of the digit which is the coefficient of $m_i$,
the metallic component of~$a$, $b$ respectively. We assume to read the places
from right to left, starting from place~0. The second operation consists in
constructing in a table~{\bf [$c$]}, the {\bf complement to $m_k$}, {\it i.e.}
where $c$ is defined by \hbox{$c+b=m_{k}$},
$k$ being the length of {\bf [$b$]}. Let us write \hbox{$a=\alpha_k m_k+a_1$}
and \hbox{$b=\beta_k m_k+b_1$}. We may assume $\alpha_k>\beta_k$: otherwise, $a-b$
is not changed if we remove from both numbers the equal leading digits until we find
the unequal ones, as we assume \hbox{$a>b$}. Then,
\hbox{$a-b=(\alpha_k-\beta_k)m_k+a_1-b_1= ((\alpha_k$$-$$1)-\beta_k) m_k 
+ (m_k-b_1)+a_1$}.
Note that \hbox{$\alpha_k$$-$$1\geq\beta_k$}. Accordingly, provided we may define the
complement of~$b_1$ to~$m_k$, we reduced the subtraction to three additions.

First, we define the comparison algorithm, again, assuming that the numbers
are given in metallic representations which are free from forbidden pattern.

\vtop{
\begin{algo}\label{acomparemc}
Comparison of $a$ and $b$ from {\bf [$a$]} and {\bf [$b$]}.
We assume that the lengths of the corresponding tables are equal. The answer is
given by the variables {\rm bigger} and {\rm smaller}.
\end{algo}
\vskip -8pt
\ligne{\hfill
\vtop{\leftskip 0pt\hsize=225pt
\hrule height 0.3 pt depth 0.3pt width \hsize
\vskip 5pt
\ligne{bigger := \faux; smaller := \faux;\hfill}
\ligne{\hskip 10pt \ffor{} i \iin{} [0..$k$] {\bf in reverse}\hfill}
\ligne{\hskip 30pt \lloop{} \iff{} {\bf [$a$]}$(i) \not=$ [$b$]$(i)$\hfill}
\ligne{\hskip 30pt \hskip 40pt \tthen{} \iff{} {\bf [$a$]}$(i) >$ [$b$]$(i)$\hfill}
\ligne{\hskip 30pt \hskip 40pt \hskip 40pt \tthen{} bigger := \vrai;\hfill}
\ligne{\hskip 30pt \hskip 40pt \hskip 40pt \eelse{} smaller := \vrai;\hfill}
\ligne{\hskip 30pt \hskip 40pt \hskip 25pt \endif;\hfill}
\ligne{\hskip 30pt \hskip 40pt \hskip 25pt {\bf exit};\hfill}
\ligne{\hskip 30pt \hskip 25pt \endif;\hfill}
\ligne{\hskip 30pt \endloop;\hfill}
\vskip 5pt
\hrule height 0.3 pt depth 0.3pt width \hsize
}
\hfill}
}
\vskip 5pt
\vskip 5pt
In order to write the algorithm computing the complement to $m_k$,we introduce
an additional convention. If {\bf [$a$]} is the metallic code of a number,
\hbox{\bf [$a$]($u..v$)} is a sub-word of~$[a]$  going from the place~$u$ to the 
place~$v$, assuming that \hbox{$u\leq v$}. In the algorithm, \hbox{\bf [$a$]($u..v$)}
can be a value and it can be addressed a value by affectation or by an operation.
The addition described by Algorithm~\ref{aaddmetal} is denoted by $\oplus$.

The first idea is to take {\bf dc$^{k-2}$d} instead of {\bf 1$^k$} in order to 
represent $m_k$. We again denote {\bf dc$^{k-2}$d} by {\bf [$a$]}. We use this trick 
as far as for most digits {\bf [$b$]$(i)$}, they are not greater than~\ccc. But for 
a few of them, we may have \hbox{\bf [$b$]$(i) >$ [$a$]$(i)$}. When it is the case,
we say that we have an {\bf inversion} at~$i$. Note that \hbox{$i<k$} as 
\hbox{\bf [$a$]$(k)$ $=$ d}. Assume that the first digit of~{\bf [$b$]}
is less than~\ddd. In that case, we split {\bf [$a$]} as follows:
\hbox{\bf [$a$]$(0..k) =$ [$a$]$(i$$+$$1..k)$$+$[$a$]$(0..i)$}, and we perform
the following transformations. Taking an auxiliary table {\it aux} of size~$k$+1 
initialized with \zzz's, we define \hbox{\bf {\it aux}$(0..i) =$ [$a$]$(0..i)$},
then we set \hbox{\bf [$a$]$(i$$+$$1)$} to \hbox{\bf [$a$]$(i$$+$$1)$$-$$1$} and
we set \hbox{\bf [$a$]$(0..i) =$ dc$^{i-1}$d}. Call these changes the {\bf lifting}.
Note that a lifting leaves {\bf [$b$]} unchanged but it changes the value of~$a$
by subtracting {\bf ([($a$)]$(0..i)$)}. It is the reason why we saved the replaced 
digits of~{\bf [$a$]} in {\it aux}. However, the result
of the lifting will eventually put in {\bf [$a$]} digits which are never
less than those of {\bf [$b$]}, so that the subtraction can be performed digit by digit.
Now, as we changed the value of~$a$, we have to add the value of the saved digits
to the result we have obtained. However, there may be several inversions in~$b$.
Moreover, the lifting may raise a new inversion. Nonetheless,
the inversion will eventually lead to a new value of~{\bf [$a$]} whose digits
are not less than the corresponding digits of~{\bf [$b$]}.
Before turning to the situation of several liftings, note that after the first
lifting, we have to add the digits we have to save to the current value of~{\it aux}:
we should not forget the former values, as each lifting consists in subtracting a
value from~{\bf [$a$]} which must be added to the final subtracting digit by digit.

\vtop{
\begin{algo}\label{acomplement}
We assume that \hbox{\bf [$a$]$(0..k) =$ dc$^{k-1}$d} that the length of
\hbox{\bf [$b$]} is $k$$+$$1$. The result is in {\bf [$c$]}.
\end{algo}
\vskip-8pt
\ligne{\hfill
\vtop{\leftskip 0pt\hsize=275pt
\hrule height 0.3 pt depth 0.3pt width \hsize
\vskip 5pt
\ligne{aux$(0..k)$ := \zzz$^{k}+1$; inversion := \faux; \hfill}
\ligne{\wwhile{} \nnot{} inversion\hfill}
\ligne{\hskip 10pt \lloop{} \ffor{} i \iin{} [0..$k$] {\bf in reverse}\hfill}
\ligne{\hskip 10pt\hskip 40pt \lloop{} \iff{} \bbb$(i) >$ \aa$(i)$\hfill}
\ligne{\hskip 10pt\hskip 40pt \hskip 30pt \tthen{} 
inversion := \vrai; place := $i$; {\bf exit}; \hfill}
\ligne{\hskip 10pt\hskip 40pt \hskip 30pt \eelse{} 
inversion := \faux; \hfill}
\ligne{\hskip 10pt \hskip 40pt \hskip 25pt \endif;\hfill}
\ligne{\hskip 10pt \hskip 40pt \endloop;\hfill}
\ligne{\hskip 10pt \hskip 30pt \iff{} inversion \hfill}
\ligne{\hskip 10pt \hskip 35pt \tthen{} 
aux(place+1$..k$) := aux $\oplus$ {\bf [$a$]}(place+1$..k$);\hfill}
\ligne{\hskip 10pt \hskip 35pt \hskip 25pt 
{\bf [$a$]}(place+1) := {\bf [$a$]}(place+1) $-$ 1;\hfill}
\ligne{\hskip 10pt \hskip 35pt \hskip 25pt 
{\bf [$a$]}(0..place) := {\bf dc$^{{\rm place}-1}$d};\hfill}
\ligne{\hskip 10pt \hskip 30pt \endif;\hfill}
\ligne{\hskip 10pt \endloop;\hfill}
\ligne{\ffor{} $i$ \iin{} [0..$k$] \hfill}
\ligne{\hskip 20pt \lloop{} {\bf [$c$]$(i)$ := [$a$]$(i) -$ [$b$]$(i)$};\hfill}
\ligne{\endloop;\hfill}
\ligne{{\bf [$c$] := [$c$] $\oplus$} aux;\hfill}
\vskip 5pt
\hrule height 0.3 pt depth 0.3pt width \hsize
}
\hfill}
}
\vskip 10pt
Presently, let us look closer at the lifting of an inversion. We defined the place~$i$ 
where the inversion occurs and acted on the digit at~$i$+1 and on the part $(0..i)$ 
of~{\bf [$a$]}. The change \hbox{\bf [$a$]$(i$$+$$1)$ $:=$ [$a$]$(i$$+$$1)$$-$$1$}
entails an inversion in the case when \hbox{\bf [$a$]$(i$$+$$1)$ = [$b$]$(i$$+$$1)$}
before the lifting. Remember that, by definition of the inversion,
\hbox{\bf [$a$]$(i$$+$$1)$ $\geq$ [$b$]$(i$$+$$1)$} before the lifting. In that
case, a new inversion occurs at~$i$+1. Now, we claim that to the left of the place of the
first lifting and in between an occurrence of~\ddd{} in~{\bf [$a$]} to the next one
to the right, we may always assume that there is an index~$j$ such that
\hbox{\bf [$a$]$(j)$ $>$ [$b$]$(j)$}. It is plain for two occurrences of~\ddd{}
inside {\bf [$b$]} as there is no forbidden pattern in {\bf [$b$]}. Consider
the case of the first lifting. It may happen that \hbox{\bf [$a$]$(j)$ $=$ c} for
all \hbox{$j\in[i$+1$..k]$}. In that case, the first lifting raises an inversion
at~$i$+1. A new lifting raises a new one at~$i$+2 raising again the initial 
inversion at~$i$. This process is repeated until the inversion occurs at~$k$$-$1.
Now, at~$k$$-$1 the lifting raises no more inversion and the leftmost \ddd{} 
occurs now in~$k$$-$1. The inversion at~$i$ is still present, so that the process is
repeated until the lifting occurs at $k$$-$2, placing \ddd{} at this place.
We can see that the process is repeated until \ddd{} is placed at $i$+1. At that
moment, the lifting replaces \ddd{} by~\ccc, so that presently
\hbox{\bf [$a$]($i$$+$$1) =$ [$b$]$(i$$+$$1)$}. Accordingly, all digits of {\bf [$a$]}
and {\bf [$b$]} from the place~$k$ to the place~$i$+1 are equal so that we may
forget them and we have \hbox{\bf [$a$]$(i)=$ d} while \hbox{\bf [$b$]$(i) =$ d} too,
so that those digits are equal too. The proof of the correctness of
Algorithm~\ref{acomplement} is completed. \hfill $\Box$

We conclude that subsection by looking at two additional algorithms: one for 
incrementing the metallic code of a number and the other for decrementing it.
For these operations, we need an algorithm which transforms a metallic representation
of~$n$ into its metallic code~{\bf[$n$]} which does not contain any occurrence
of the forbidden patterns. We start with that algorithm:

\vtop{
\begin{algo}\label{alibre}
The representation is given in {\bf a}, {\bf a}$(i)$ being its $i^{\rm th}$ digit.
\end{algo}
\vskip -8pt
\ligne{\hfill
\vtop{\leftskip 0pt\hsize=300pt
\hrule height 0.3pt depth 0.3pt width \hsize
\vskip 5pt
\ligne{warning := false; $i$ := 0;\hfill}
\ligne{\wwhile{} i $\leq k$\hfill}
\ligne{\hskip 10pt \lloop{} \iff{} \aa$(i) =$ \ddd\hfill}
\ligne{\hskip 10pt \hskip 30pt \tthen{} \iff{} \nnot{} warning\hfill}
\ligne{\hskip 10pt \hskip 30pt \hskip 30pt \tthen{}
warning := \vrai; init := $i$;\hfill}
\ligne{\hskip 10pt \hskip 30pt \hskip 30pt \hskip 27.5pt
$i$ := $i$+1;\hfill}
\ligne{\hskip 10pt \hskip 30pt \hskip 30pt \eelse{}
\iff{} \aa$(i) =$ \ddd\hfill}
\ligne{\hskip 10pt \hskip 30pt \hskip 30pt \hskip 30pt\tthen{}
final := $i$;\hfill} 
\ligne{\hskip 10pt \hskip 30pt \hskip 30pt \hskip 30pt\hskip 27.5pt
\ffor{} $j$ \iin{} [init..final] \hfill}
\ligne{\hskip 10pt \hskip 30pt \hskip 30pt \hskip 30pt\hskip 27.5pt\hskip 5pt
\lloop{} \aa$(i)$ := 0; \endloop; \hfill}
\ligne{\hskip 10pt \hskip 30pt \hskip 30pt \hskip 30pt\hskip 27.5pt
\iff{} init $>$ 0\hfill}
\ligne{\hskip 10pt \hskip 30pt \hskip 30pt \hskip 30pt\hskip 27.5pt\hskip 5pt
\tthen{} \aa{}(init$-$1) := \aa{}(init$-$1)+1;\hfill}
\ligne{\hskip 10pt \hskip 30pt \hskip 30pt \hskip 30pt\hskip 27.5pt
\endif;\hfill}
\ligne{\hskip 10pt \hskip 30pt \hskip 30pt \hskip 30pt\hskip 27.5pt
\aa(init+1) := \aa(init+1)+1;\hfill}
\ligne{\hskip 10pt \hskip 30pt \hskip 30pt \hskip 30pt\hskip 27.5pt
$i$ := 0; warning := \faux; exit;\hfill}
\ligne{\hskip 10pt \hskip 30pt \hskip 30pt \hskip 30pt\eelse{}
\iff{} \aa$(i) <$ \ccc\hfill}
\ligne{\hskip 10pt \hskip 30pt \hskip 30pt \hskip 30pt\hskip 27.5pt\tthen{}
warning := \faux;\hfill}
\ligne{\hskip 10pt \hskip 30pt \hskip 30pt \hskip 30pt\hskip 20pt
\endif;\hfill}
\ligne{\hskip 10pt \hskip 30pt \hskip 30pt \hskip 30pt\hskip 20pt
$i$ := $i$+1;\hfill}
\ligne{\hskip 10pt \hskip 30pt \hskip 30pt \hskip 20pt
\endif;\hfill}
\ligne{\hskip 10pt \hskip 30pt \hskip 27.5pt 
\endif;\hfill} 
\ligne{\hskip 10pt \hskip 30pt \eelse{} \iff{} \aa$(i) <$ \ccc\hfill}
\ligne{\hskip 10pt \hskip 30pt \hskip 30pt \tthen{}
warning := \faux;\hfill}
\ligne{\hskip 10pt \hskip 30pt \hskip 22.5pt
\endif;\hfill}
\ligne{\hskip 10pt \hskip 30pt \hskip 22.5pt
$i$ := $i$+1;\hfill}
\ligne{\hskip 10pt \hskip 25pt \endif;\hfill} 
\ligne{\hskip 10pt \endloop;\hfill} 
\vskip 5pt
\hrule height 0.3pt depth 0.3pt width \hsize
}
\hfill}
}
\vskip 10pt
\noindent
Note that Algorithm~\ref{alibre} eliminates all forbidden pattern from the metallic
representation of~$n$.

Thanks to that algorithm, we assume that we consider the metallic code of~$n$, {\it i.e.}
the metallic representation of the number which is free of forbidden pattern. Denoting
Algorithm~\ref{aaddmetal} by $\oplus$ and the subtraction by~$\ominus$,
we might define the operation of incrementing {\bf [$n$]} by \hbox{\bf [$n$]$\oplus$\uu}
and the operation of decrementing the same code by \hbox{\bf [$n$]$\ominus$\uu}. 
However, for those particular operations, it is possible to provide simpler algorithms,
see Algorithm~\ref{amincr} and Algorithm~\ref{amdecr}, below.

\vtop{
\begin{algo}\label{amincr}
Algorithm for writing \hbox{\bf [$n$$+$$1$]} knowing \aa{} $=$ {\bf [$\nu$]}.
\end{algo}
\vskip-8pt
\ligne{\hfill
\vtop{\leftskip 0pt\hsize=230pt
\hrule height 0.3pt depth 0.3pt width \hsize
\vskip 5pt
\ligne{\hskip 10pt \iff{} \aa$(0) =$ \ddd\hfill}
\ligne{\hskip 10pt \hskip 5pt \tthen{} \aa$(0)$ := $0$;
\aa$(1)$ := \aa$(1)$+1;\hfill}
\ligne{\hskip 10pt \hskip 5pt \eelse{}
\iff{} \aa$(0) <$ \ccc\hfill}
\ligne{\hskip 10pt \hskip 30pt \tthen{} \aa$(0)$ := \aa$(0)$+1;\hfill}
\ligne{\hskip 10pt \hskip 30pt \eelse{}
$i$ := $0$;\hfill} 
\ligne{\hskip 10pt \hskip 30pt \hskip 25pt \wwhile{} \aa$(i)$ = \ccc\hfill}
\ligne{\hskip 10pt \hskip 30pt \hskip 25pt \hskip 10pt
\lloop{} $(i)$ := $i$+1; \endloop;\hfill}
\ligne{\hskip 10pt \hskip 30pt \hskip 25pt \iff{} \aa$(i)$ $<$ \ddd\hfill}
\ligne{\hskip 10pt \hskip 30pt \hskip 25pt \hskip 5pt \tthen{}
\aa$(0)$ = \ddd;\hfill}
\ligne{\hskip 10pt \hskip 30pt \hskip 25pt \hskip 5pt \eelse{}
\ffor{} $j$ \iin{} [0..$i$] \hfill}
\ligne{\hskip 10pt \hskip 30pt \hskip 25pt \hskip 5pt \hskip 30pt
\lloop{} \aa$(j)$ := 0; \endloop;\hfill}
\ligne{\hskip 10pt \hskip 30pt \hskip 25pt \hskip 5pt \hskip 22.5pt
\aa$(i$+$1)$ := \aa$(i$+$1)+1$;\hfill}
\ligne{\hskip 10pt \hskip 30pt \hskip 25pt \endif;\hfill}
\ligne{\hskip 10pt \hskip 27.5pt \endif;\hfill}
\ligne{\hskip 10pt \endif;\hfill}
\vskip 5pt
\hrule height 0.3pt depth 0.3pt width \hsize
}
\hfill}
}
\vskip 15pt
\vtop{
\begin{algo}\label{amdecr}
Algorithm for writing \hbox{\bf [$n$$-$$1$]} knowing \aa{} $=$ {\bf [$\nu$]}.
\end{algo}
\vskip-8pt
\ligne{\hfill
\vtop{\leftskip 0pt\hsize=220pt
\hrule height 0.3pt depth 0.3pt width \hsize
\vskip 5pt
\ligne{\hskip 10pt $i$ := $0$;\hfill} 
\ligne{\hskip 10pt \wwhile{} \aa$(i)$ = 0 \lloop{} $(i)$ := $i$+1; \endloop;\hfill}
\ligne{\hskip 10pt \aa$(i)$ := \aa$(i)$$-$1;\hfill} 
\ligne{\hskip 10pt \ffor{} $j$ \iin{} [0..$i$$-$2] \lloop{} \aa$(j)$ := \ccc; 
\endloop;\hfill}
\ligne{\hskip 10pt \iff{} $i$ $\geq$ 1\hfill}
\ligne{\hskip 10pt \hskip 5pt \tthen{}
\aa$(i$$-$$1)$ = \ddd;\hfill}
\ligne{\hskip 10pt \endif;\hfill}
\vskip 5pt
\hrule height 0.3pt depth 0.3pt width \hsize
}
\hfill}
}
\vskip 10pt
Note that \aa{} is supposed to be a metallic code and that, accordingly, the
result is a metallic code too.
From these algorithms, we can see that successively incrementing {\bf [$n$]},
the change behaves as if {\bf [$n$]} were written in basis~$p$$-$2 until
a pattern \hbox{\ddd\ccc$^\ast$\ddd} occurs as a suffix of {\bf [$n$$+$$m$]}.
Then, the pattern is replaced by the same number of~\zzz's as its length
and by adding a carry~1 to the rest of the representation. We shall also use that
property in the proofs of the properties which will be reported in 
Sections~\ref{white_metal} and~\ref{black_metal}.

\section{Properties of the white metallic tree}\label{white_metal}

After defining the basic operations on the metallic code we shall use in this
section and in the next ones, we look at the metallic codes of the numbers
we met about the metallic trees, namely $m_n$, $b_n$,$M_n$ and $B_n$.
From Theorem~\ref{theadlevelwb} and the relations~$(5)$, we have:

\begin{lemm}~\label{lcodesLwb}
In the white metallic tree, the number of nodes on the level~$n$ is $m_n$
and \hbox{\bf [$m_n$] $=$ 10$^k$}. The rightmost node on that level is~$M_n$
and \hbox{\bf [$M_n$] = 1$^{k+1}$}.

In the black metallic code, the number of nodes on the level~$n$ is $b_n$
and \hbox{\bf [$b_n$] $=$ c$^{n-1}$d}. As \hbox{$B_n=m_n$}, we conclude
that \hbox{\bf [$B_n$] $=$ 10$^k$}.
\end{lemm}

\noindent
Proof. The metallic codes of~$m_n$, hence for~$B_n$ are trivially computed.
For {\bf [$b_n$]}, we apply the proof of the correctness of Algorithm~\ref{acomplement}.
Indeed, it appears from \hbox{$b_{n+1} = m_{n+1}-m_n$} that \hbox{\bf [$b_{n+1}$]}
is the complement to~$m_{n+1}$ of~$m_n$. Writing {\bf [$b_{n+1}$]} as
{\bf dc$^{n-1}$d} and {\bf [$m_n$]} as {\bf 10$^n$}, we immediately get that
\hbox{\bf [$b_{n+1}$] $=$ c$^n$d}, as far as all digits of {\bf [$m_n$]} or not
greater than the corresponding ones of~{\bf dc$^{n-1}$d}.\hfill $\Box$

Consider the white metallic tree $\cal W$. We can prove the following result:

\begin{lemm}\label{lmwdecomp1}
Denote by ${\cal B}_{\numd,n}$, ${\cal W}_{\numt,n}$, ..., ${\cal W}_{\ddd,n}$,
${\cal W}_{\uu\zzz,n}$ and ${\cal W}_{\uu\uu,n}$ the metallic sub-trees of $\cal W$ of 
height~$n$ rooted at 
the nodes \numd, \numt, ..., \ddd, \uu\zz{} and \uu\uu{} the sons of the root \uu.
Denote by $\rho_{\numd,n}$, $\rho_{\numt,n}$, ..., $\rho_{\ddd,n}$ and  
$\rho_{\uu\zzz,n}$ and $\rho_{\uu\uu,n}$ the rightmost node of the respective sub-trees 
on the level~$n$$+$$1$ of $\cal W$. We get that:
\vskip 5pt
\ligne{\hfill
$\vcenter{\vtop{\leftskip 0pt\hsize=220pt
	\ligne{{\bf [$\rho_{\numd,n}$]} $=$ \numd\zzz\uu$^{n-1}$, 
	{\bf [$\rho_{\numt,n}$]} $=$ \numt\zzz\uu$^{n-1}$,\hfill} 
	\ligne{\hfill ...,\hfill}
	\ligne{{\bf [$\rho_{\ddd,n}$]} $=$ \ddd\zzz\uu$^{n-1}$, 
	{\bf [$\rho_{\uu\zzz,n}$]} $=$ \uu\zzz\uu$^n$, 
	{\bf [$\rho_{\uu\uu,n}$]} $=$ \uu$^{n+2}$,\hfill} 
}}$
\hfill $(12)$\hskip 10pt}
\end{lemm}

\noindent
Proof. From Lemma~\ref{lcodesLwb}, we know that the rightmost son of 
${\cal L}_{n,\cal W}$ is $M_n$ and that \hbox{\bf [$M_n$] $=$ 1$^{n+1}$}.
We also know that \hbox{\bf [$b_n$] $=$ c$^{n-1}$d}. Accordingly,
we obtain that \hbox{{\bf [$\rho_{\numd,n}$]} = \uu$^{n+1}\oplus \ccc^{n-1}\ddd$}, 
which can be computed by Algorithm~\ref{aaddmetal}. The computation gives rise to:
\vskip 5pt
\def\regul #1 #2 #3 #4 {%
\vtop{\leftskip 0pt\hsize=50pt
\ligne{\hfill #1 \hfill}
\ligne{\hfill #2 \hfill}
\ligne{\vrule height 0.3pt depth 0.3pt width \hsize}
\ligne{\hfill #3 \hfill}
\vskip-5pt
\ligne{\hfill #4 \hfill}
}
}
\ligne{\hfill
\regul {\uu\uu...\uu\uu} {\zzz\ccc...\ccc\ddd} {\uu\ddd...\ddd\zzz}
{\footnotesize \hskip 17pt \uu\hskip 5pt\uu}
\hskip 20pt
\regul {\uu\ddd...\ddd\zzz}
{\hskip 17pt \uu\hskip 5pt\uu}
{\uu\ddd...\zzz\uu}
{\footnotesize \hskip 7pt \uu\hskip 5pt\uu}
\hskip 20pt
...
\hskip 20pt
\regul {\uu\ddd...\uu\uu}
{\hskip-15pt\uu\hskip 5pt\uu}
{\numd\zzz...\uu\uu}
{}
\hfill}
\vskip 5pt
\noindent
which prove the result for $\rho_{\numd,n}$. Now, for the following nodes, we simply
append $m_n$ at each step and, as \hbox{\bf [$m_n$] $=$ \uu\zzz$^n$}, we get the
result of the lemma up to $\rho_{\ddd,n}$ being included. For the two rightmost sub-tree,
we have that \hbox{{\bf [$\rho_{\uu\zzz,n}$]} = {\bf [$\rho_{\ddd,n}$]} $\oplus~m_n$} and
that \hbox{{\bf [$\rho_{\uu\uu,n}$]} = {\bf [$\rho_{\uu\zzz,n}$]} $\oplus~m_n$}, 
{\it i.e.}:
\vskip 5pt
\ligne{\hfill
\regul {\zzz\ddd\zzz...\uu\uu} {\zzz\uu\zzz...\zzz\zzz} {\zzz\zzz\zzz...\uu\uu}
{\footnotesize\hskip-20pt\uu\hskip 5pt\uu} \hskip 40pt
\regul {\uu\zzz\uu...\uu\uu} {\zzz\uu\zzz...\zzz\zzz} {\uu\uu\uu...\uu\uu}
{}
\hfill}
\vskip 5pt
The proof of the lemma is completed.\hfill $\Box$
   
   We can now state the following property:

\begin{thm}\label{tmpreferred} 
In a white metallic tree, for any node~$\nu$ we have that among its sons
a single one has $[\nu]0$ as its metallic code, which is called the 
{\bf preferred son} of~$\nu$. In order to find out which son of~$\nu$ is its
preferred one, we distinguish two kinds of white nodes : {\bf w}$_\ell$ 
and {\bf w}$_r$. In a black node and in a {\bf w}$_\ell$-node, 
its preferred son is its last one. In a {\bf w}$_r$-node, its preferred son
is its penultimate one. Moreover, the nodes obey the following rules :
\vskip 5pt
\ligne{\bf\hfill b $\rightarrow$ bw$_\ell^{p-5}$w$_r,$ 
w$_\ell$ $\rightarrow$ bw$_\ell^{p-4}$w$_r,$
	w$_r$ $\rightarrow$ bw$_\ell^{p-5}$w$_r$w$_r.$\hfill$(13)$\hskip 10pt}
\vskip 5pt
\noindent
Denote $p$$-$$3$ by~\ddd, $p$$-$$4$ by~\ccc, $p$$-$$5$ by \eee{} and by 
\aa{} any digit
in \hbox{$[$\numd..\ddd$]$}. We also have the following rules on the signatures 
where type and signature are associated on the left hand-side of the rule:
\vskip 5pt
\ligne{\hfill 
$\vcenter{\vtop{\leftskip 0pt\hsize=300pt\bf
\ligne{\hfill
	\bbb\uu$,$\bbb\numd{} $\rightarrow$ \bbb\numd$($\www\aa$)^{p-5}$\www\zzz$,$
	\www\aa{} $\rightarrow$ \bbb\uu$($\www\aa$)^{p-4}$\www\zzz$,$ 
\hfill}
\ligne{\hfill
	\www\zzz{} $\rightarrow$ \bbb\uu$($\www\aa$)^{p-6}$\www\ccc\www\zzz\www\uu$,$
	\www\uu{} $\rightarrow$ \bbb\numd$($\www\aa$)^{p-6}$\www\ddd\www\zzz\www\uu$,$ 
\hfill}
}}$
\hfill$(14)$\hskip 10pt}
\vskip 5pt
\noindent
so that \www\aa{} is a \www$_\ell$-node while \www\zzz{} and \www\uu{}
are \www$_r$-nodes.
At last, for any non-negative integer~$k$, $m_{k+1}$ is the preferred son
of~$m_k$. 
\end{thm}

\noindent
Proof.
Figure~\ref{fmetalblanc} illustrates the properties stated in the theorem. 
In the figure, $p=9$ and we did not represent all the sons of a node for clarity reasons.
Enough nodes are documented which allows the reader to note that the properties 
stated in the theorem are observed. For the nodes which are documented, the number is
written in red and displayed as usual. The metallic code is written in purple
and vertically under the node.
In the figure, we represent black nodes in red colour.
White nodes are represented in blue for {\bf w$_\ell$}-nodes,
in green for {\bf w$_r$}-nodes. Moreover, green nodes which are also preferred sons
of their father are represented by a green disc with a red border. 

\vskip 10pt
\vtop{
\ligne{\hfill
\includegraphics[scale=0.35]{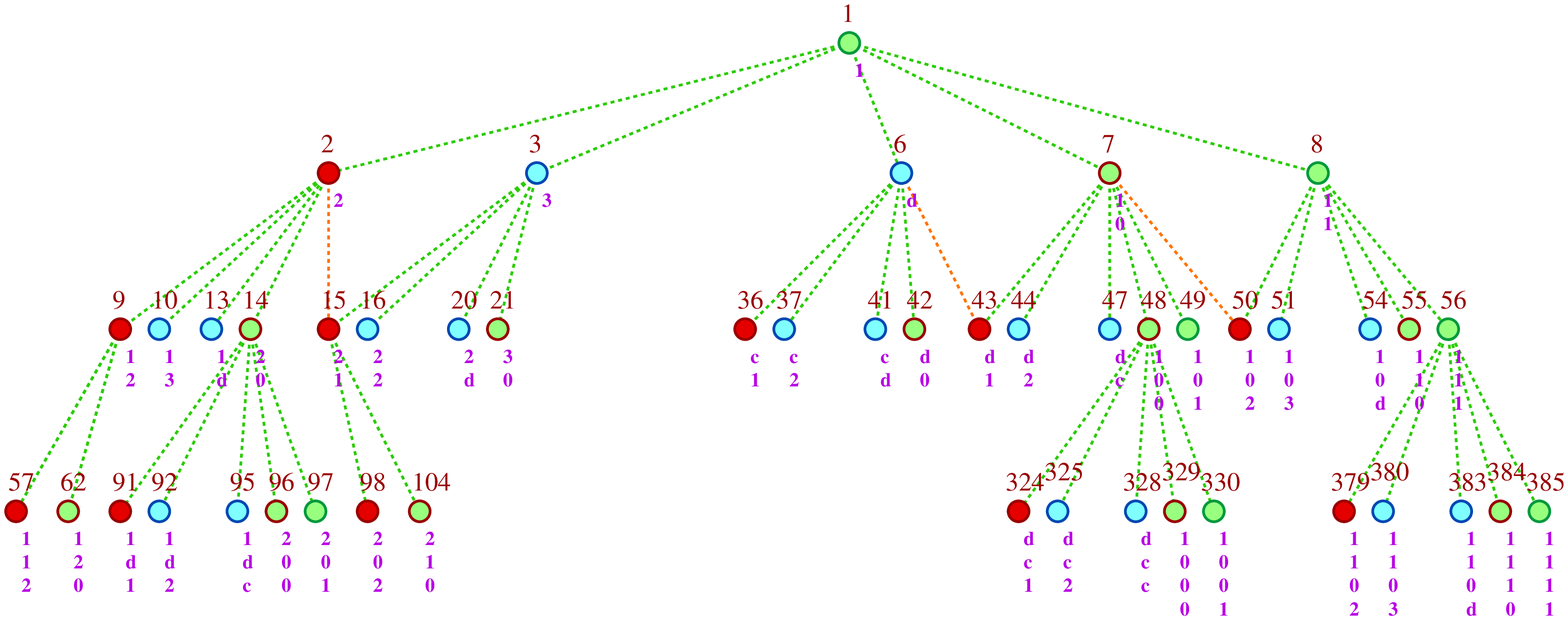}
\hfill}
\vspace{-10pt}
\ligne{\hfill
\vtop{\leftskip 0pt\parindent 0pt\hsize=300pt
\begin{fig}\label{fmetalblanc}
\leurre
The white metallic tree. Partial representation of the first three levels of the tree
when $p=9$
with the conventions mentioned in the text.
\end{fig}
}
\hfill}
}
\vskip 5pt
The figure is intended to help us to perform the proof of Theorem~\ref{tmpreferred}.

In our proof and later on, we denote by ${\cal W}_{\nu}$ the white metallic tree rooted 
at the white node~$\nu$,
and by ${\cal B}_{\beta}$ the black metallic tree rooted
at the black node~$\beta$. Later, in the index, the occurrence of~$n$ will indicate 
that we consider the sub-tree of height~$n$ issued from the same root.

First, note that although Lemma~\ref{lmwdecomp1} seems to give global information and
although it is precise on the extremal branches of each sub-tree only, it can be used
for more local information, provided we know the distance of a node~$\nu$ from the
borders of the sub-tree containing~$\nu$ we consider.

As an example, consider the location of the nodes $m_n$. 
We have seen that Lemma~\ref{lmwdecomp1} proved that 
\hbox{\bf [$\rho_{\uu\uu,n}$] $=$ \uu$^{n+1}$} and that 
\hbox{\bf [$m_n$] $=$ 10$^n$}. Let us look at what it means on levels~1, 2 and~3.
On level~1, {\bf [$m_1$] $=$ \uu\zzz}, so that it is the penultimate son of the root~1.
On level~2, {\bf [$m_2$] $=$ \uu\zzz\zzz}, so that its distance from~$M_2$,
the rightmost node of level~2 is {\bf \uu\uu} which can be split into
\hbox{$m_1$+$m_0$}. Now, $m_1$ is the distance from $\rho_{\uu\uu,1}$ to
$\rho_{\uu\zzz,1}$, so that \hbox{$m_0=1$} is the distance from $m_2$ to
$\rho_{\uu\zzz,1}$: indeed, the level~1 of ${\cal W}_{\uu\zzz}$ is on the level~2 of 
${\cal W}$. Accordingly, $m_2$ is the penultimate son of the root~$m_1$ of
${\cal W}_{\uu\zzz}$. The same decomposition for \hbox{$M_3-m_3$} shows us that
inside ${\cal W}_{\uu\zzz}$ at which we arrive thanks to~$m_2$ which is the
distance from~$\rho_{\uu\uu,2}$ to $\rho_{\uu\zzz,2}$. Define $\pi_1$ to 
be $\rho_{\uu\uu,1}$, the root of ${\cal W}_{\uu\uu}$. Then, define $\pi_2$ to be the
root of the sub-tree of~${\cal W}_{\uu\zzz}$ rooted at $m_1$+1. By $m_1$, we arrive
from the rightmost node of ${\cal L}_{{\cal W}_{\pi_2},1}$ to a node~$\pi_3$
on the level~2 of~${\cal W}_{\uu\zzz}$ which is at the distance~$m_0$ from~$m_3$.
Accordingly, $m_3\in{\cal W}_{m_2}$ and $m_3$ is the penultimate son of~$m_2$.

By induction, denote~$\pi_n$ the node of the level~$n$ of~$\cal W$ which is in between
$m_n$ and $M_n$ at the distance~$m_0$ from~$m_n$. Also, assume that the distance
of~$m_n$ from $\rho_{\uu\uu,n-1}$ is $m_{n-1}$ as already known allows us to cross
the sub-trees ${\cal W}_{\pi_1}$, ..., ${\cal W}_{\pi_{n}}$ whose heights are
$n$$-$1, ..., 0 respectively. On the level~$n$+1 of~$\cal W$, when we go 
from~$\rho_{\uu\uu,n}$ to~$m_{n+1}$, the level of the crossed nodes is
$n$ in ${\cal W}_{\pi_1}$, ..., 1{} in ${\cal W}_{\pi_n}$ respectively and the number 
of nodes which are crossed is $m_n$, ..., $m_1$ respectively. Accordingly, after crossing
$\displaystyle{\sum\limits_{i=1}^n m_i}$ nodes from~$\rho_{\uu\uu,n}$, we arrive
to the rightmost son of~$\pi_n$: it is the level~1 of ${\cal W}_{m_n}$. It is $\pi_{n+1}$
and we remain with the crossing of that node in order to reach $m_{n+1}$, consuming
$m_0$, so that we crossed \hbox{$M_{n+1}-m_n=M_n$} nodes. This proves that $m_{n+1}$
is the penultimate son of~$m_n$ and we proved the induction hypothesis. Accordingly,
the last property stated in Theorem~\ref{tmpreferred} is proved.

%%%%%
Let us prove the other assertions of the theorem. Applying Algorithms~\ref{amincr}
and~\ref{amdecr}, we assume that those properties are true for all nodes whose number
is at most~$\nu$: it means that it is true for all nodes of ${\cal W}_n$, where $n$+1
is the level of~$\nu$ and, on the level~$n$+1 for all nodes whose number~$\mu$ is less
than~$\nu$. The computations of Lemma~\ref{lmwdecomp1} show us that the relations~$(13)$ 
and~$(14)$ are true for the nodes of level~1: they are immediate consequences of~$(12)$.

Assume that $\nu$ is the leftmost node on the level~$n$+1. Its number is $M_n$+1, 
so that \hbox{\bf [$\nu$] $=$ \uu$^n$\numd} as deduced from Lemma~\ref{lmwdecomp1}.
Let~$\sigma$ be the leftmost node of~$\nu$. From the same lemma,
\hbox{\bf [$\sigma$] $=$ \uu$^{n+1}$\numd}, so that the repetition of 
Algorithm~\ref{amincr} shows us that the son signature of~$\nu$ is
\hbox{\bf 23..d0} so that we have the rule 
\hbox{\numd{} $\rightarrow$ \numd\numt..\ddd\zzz}.

Next, we display the proof in Table~\ref{tmwsons} which concentrates
the computations performed by the iterated application of Algorithm~\ref{amincr}.
Denote by~$\sigma_\ell(\nu)$, $\sigma_r(\nu)$ the leftmost,
rightmost son respectively of~$\nu$. From the definitions we easily get:
\vskip 5pt
\ligne{\hfill
$\sigma_\ell(\nu$+$1) = \sigma_r(\nu)+1$ and
$\sgn{}(\sigma_\ell(\nu$+$1) = \sgn{}(\nu)\oplus\uu$.
\hfill $(15)$\hskip 10pt}
\vskip 5pt
\noindent
To better understand the construction of the table, we make use of 
Table~\ref{tsignatures} which give the possible sons signatures of a node
assuming the signature of its leftmost son. By induction hypothesis,
we assume that we have
\hbox{\sgn{}$(\sigma_\ell(\nu))\in\{\uu, \numd\}$} for the node~$\nu$,
wheter it is black or white.
Tables~\ref{tsignatures} and~\ref{tmwsons} also take into account that a
black node has (\ddd) sons and that a white one has (\uu\zzz) of them. 
Note that \ccc{} is followed by~\zzz{} if and only if a suffix \hbox{\ddd\ccc$^\ast$} 
occurs in \hbox{\bf [$\nu$$-$1]}.
When going from the sons of the node~$\nu$
to those of the node~$\nu$+1, we shall use the following remark:
we also shall consider $\pi_r(\nu)$ the penultimate son of~$\nu$.

\def\fnb #1{\hbox{\footnotesize\bf #1}}
\newdimen\lelarge\lelarge=35pt
\def\lalignette #1 #2 #3 #4 #5 #6 #7 {
\ligne{
\hbox to \lelarge {#1\hfill}
\hbox to \lelarge {#2\hfill}
\hbox to \lelarge {#3\hfill}
\hbox to \lelarge {#4\hfill}
\hbox to \lelarge {#5\hfill}
\hbox to \lelarge {#6\hfill}
\hbox to \lelarge {#7\hfill}
\hfill}
}
\vtop{
\begin{tab}\label{tsignatures}
Auxiliary table for the computations of the sons signature in a white
metallic tree. We remind the reader that \eee{} satisfies
\hbox{\eee$^-+$$1$ = \ccc}. The bullet $\bullet$ indicates the part of the
table used by a black node.
\end{tab}
\vskip-8pt
\ligne{\hfill
\vtop{\leftskip 0pt\hsize 300pt
\lalignette {\fnb 1 } {\fnb 2 } {...} {\fnb {\ccc} } {\fnb {\ddd} $\bullet$} 
{\fnb {\uu\zzz} } {}
\lalignette {\zzz} {\uu} {...} {\eee} {\ccc} {\ddd} {\fnb 1 }
\lalignette {\zzz} {\uu} {...} {\eee} {\ccc} {\zzz} {\fnb 2 }
\lalignette {\uu} {\numd} {...} {\ccc} {\ddd} {\zzz} {\fnb 3 $\ast$}
\lalignette {\uu} {\numd} {...} {\ccc} {\zzz} {\uu} {\fnb 4 $\ast$}
\lalignette {\numd} {\numt} {...} {\ddd} {\zzz} {\uu} {\fnb 5 $\ast$}
\lalignette {\numd} {\numt} {...} {\zzz} {\uu} {\numd} {\fnb 4 }
}
\hfill}
}
\vskip 10pt
\def\gaa{\hbox{\bf C}}
Table~\ref{tsignatures} indicates the possible sons signatures according to the 
signature of the leftmost son of a node. The nodes are mentioned by their position
as sons of the node, from~\fnb 1 {} up to \fnb {\ddd} {} for a black node, up to 
\fnb {\uu\zzz} {}
for a white one. For the son~\fnb 1, we indicated the signatures \zzz, \uu{}
and \numd, the last two ones only occurring in the relations~$(14)$. 
The signature \zzz{} for a black node does not occur in a white metallic tree: we shall
prove that property.

\newdimen\llarge\llarge=25pt
\def\demiligne #1 #2 #3 #4 #5 #6{\footnotesize
\hbox to \llarge {\hfill#1\hfill}
\hbox to \llarge {#2\hfill}
\hbox to \llarge {#3\hfill}
\hbox to \llarge {#4\hfill}
\hbox to \llarge {#5\hfill}
\hbox to \llarge {#6\hfill}
}
\setbox150=\vtop{\leftskip 0pt\hsize=160pt
\ligne{\demiligne {\bbb} {\gaa\aa$^-$\numd} {...} {\gaa\aa$^-$\ddd} {\gaa\aa\zzz} {}
\hfill}
\ligne{\demiligne {\www} {\gaa\aa$^-$\uu} {...} {\gaa\aa$^-$\ccc} {\gaa\aa$^-$\ddd} 
{\gaa\aa\zzz}\hfill}
}
\setbox160=\vtop to \ht150{\leftskip 0pt\hsize=160pt
\vfill
\ligne{\demiligne {\www} {\gaa\aa\uu} {...} {\gaa\aa\ccc} {\gaa\aa\ddd} 
{\gaa\aa$^+$\zzz}\hfill}
\vfill
}
\vtop{
\begin{tab}\label{tmwsons}
Computation of the sons signatures in the white metallic tree. To left, $\nu$
which is supposed to observe the relations $(14)$. To right, the node $\nu$$+$$1$.
\end{tab}
\vskip-8pt
%\ligne{\hfill
%\vtop{\leftskip 0pt\hsize 330pt
\ligne{\hskip 10pt \fnb 1\hskip 90pt \gaa\aa\hskip 110pt \gaa\aa$\oplus$\uu\hfill}
\ligne{\box150 \box160 \hfill}
%}
%\hfill}
\vskip 5pt
\ligne{\hskip 10pt \fnb 2\hskip 90pt \gaa\zzz$^-$\hskip 110pt \gaa$^+$\zzz\hfill}
\ligne{
\demiligne {\www\ddd} {\gaa\ccc\uu} {...} {\gaa\ccc\ccc} {\gaa\ccc\ddd} {\gaa\ddd\zzz}
\demiligne {\www\zzz} {\gaa\ddd\uu} {...} {\gaa\ddd\ccc} {\gaa$^+$\zzz\zzz} 
{\gaa$^+$\zzz\uu} 
\hfill}
\ligne{
\demiligne {\www\ccc} {\gaa\ccc$^-$\uu} {...} {\gaa\ccc$^-$\ccc} {\gaa\ccc$^-$\ddd} 
{\gaa\ccc\zzz}
\demiligne {\www\zzz} {\gaa\ccc\uu} {...} {\gaa\ccc\ccc} {\gaa$^+$\zzz\zzz} 
{\gaa$^+$\zzz\uu} 
\hfill}
\vskip 5pt
\setbox160=\vtop{\leftskip 0pt\hsize=160pt
\ligne{\demiligne {\bbb\uu} {\gaa$^+$\zzz\numd} {...} {\gaa$^+$\zzz\ddd} 
{\gaa$^+$\uu\zzz} {} \hfill}
\ligne{\demiligne {\www\uu} {\gaa$^+$\zzz\numd} {...} {\gaa$^+$\zzz\ddd} 
{\gaa$^+$\uu\zzz} {\gaa$^+$\uu\uu} \hfill}
}
\setbox150=\vtop to \ht160{\leftskip 0pt\hsize=160pt
\vfill
\ligne{\demiligne {\www\zzz} {\gaa\ddd\uu} {...} {\gaa\ddd\ccc} {\gaa$^+$\zzz\zzz} 
{\gaa$^+$\zzz\uu} 
\hfill}
\vfill}
\ligne{\hskip 10pt \fnb 3\hskip 90pt \gaa$^+$\zzz\hskip 110pt \gaa$^+$\uu\hfill}
\ligne{\hfill\box150 \hfill\box160\hfill}
\vskip 5pt
\ligne{\hskip 10pt \fnb 4\hskip 90pt \gaa\zzz\uu\hskip 110pt \gaa\zzz\numd\hfill}
\ligne{
\demiligne {\www\uu} {\gaa$^+$\zzz\numd} {...} {\gaa$^+$\zzz\ddd} {\gaa$^+$\uu\zzz} 
{\gaa$^+$\uu\uu}
\demiligne {\bbb\numd} {\gaa$^+$\zzz\numd} {...} {\gaa$^+$\zzz\ddd} {\gaa$^+$\uu\zzz} 
{}
\hfill}
%}
%\hfill}
}
\vskip 10pt
Using $(15)$ and Table~\ref{tsignatures}, Table~\ref{tmwsons} computes
the transition from $\nu$ to~$\nu$+1 by arguing on their metallic codes only,
taking into account what we said about Algorithm~\ref{amincr} and the elimination of
a forbidden pattern. The table displays {\bf [$\nu$]} as \gaa\aa, with
\hbox{\aa{} = \sgn{}($\nu$)}. In the table, \aa$^-$ is the digit defined by
\hbox{\aa$^-$ $=$ \aa$\ominus$\uu} and \hbox{\aa$^+$ $=$ \sgn{}(\aa$\oplus$\uu)}.
We can see that when \hbox{\aa{} = \ccc}, we have \hbox{\aa$^+$ = \ddd} and we
have \hbox{\aa$^+$ = \zzz} only if \gaa{} contains a suffix of the form \ddd\ccc$^\ast$.

In part \fnb{1} of Table~\ref{tmwsons}, we have in the left-hand side as $\nu$ either
a black node or a \www\aa-node, with \hbox{\aa $\not=$ \zzz} and
\hbox{\aa $\not=$ \uu}. The node $\nu$+1 is a \www\aa-node in the case when $\nu$
is black. It is also the case for a \www\aa-node, provided that 
\hbox{\aa$^+$ $\not=$ \zzz}. The computation directly proceeds from
Table~\ref{tsignatures}: line \fnb{3} applies here.

In part \fnb{2}, we have that \hbox{\sgn{}($\nu$+1) = \zzz}. We have two cases:
the case when \gaa{} has not a suffix of the form \ddd\ccc$^\ast$, the first line
of part \fnb{2}, and the case when it does have it: the second line. In the first
line, as \gaa{} does not end with \ddd\ccc$^\ast$, we can write \gaa\ccc\ddd{}
and then, for the rightmost son, \gaa\ddd\zzz. In the second line, we
can write \gaa\ccc$^-$\ddd{} as far as \ccc\ccc$^\ast$\ddd{} is a permitted pattern.
In the right-hand side of the same line, we arrive to \gaa\ccc$^\ast$\ccc{} for
the son \fnb{\ccc}, but \gaa\ccc$^\ast$\ddd{} is a forbidden pattern so that
we get \gaa$^+$\zzz\zzz, where \hbox{\gaa$^+$ = \gaa$\oplus$\uu}.  
Note that line \fnb{4} of Table~\ref{tsignatures} applies here in both cases
and in both cases too, the rightmost son of~$\nu$+1 is a \www\uu-node, a node which 
we did not meet yet.

In part \fnb{3}, we deal with the case when $\nu$ is a \www\zzz-node. Up to now, we have 
seen two such cases: The case when such a node is the rightmost node of a black node
or of a \www\aa-node. In that case, $\nu$+1 is a black node. It is the first line of
the right-hand side of part \fnb{3}. We also met the case when $\nu$ is the penultimate
node of a \www\zzz-node. The situation is given in the second line of the right-hand side
of part \fnb{3}, an application of the line \fnb{5} of Table~\ref{tsignatures}.

At last, in part \fnb{4}, we have that $\nu$ is a \www\uu-node, a node which appeared 
in the right-hand side of part \fnb{2}. As it is the rightmost node of a white node,
$\nu$+1 is a black node. The line \fnb{5} of Table~\ref{tsignatures} again applies
but we need only the sons \fnb{1} up to \fnb{\ddd}.

The table show us that we always used the lines \fnb{3}, \fnb{4} and \fnb{5} of
Table~\ref{tsignatures} and that the other lines never appear in the new configurations.
Accordingly, $(14)$ is proved for $\nu$+1 too. The proof of Theorem~\ref{tmpreferred}
is completed.\hfill $\Box$
\vskip 5pt
Before turning to the next subsection,
we go back to a property
we noticed with the proof of~$(12)$ in Lemma~\ref{lmwdecomp1}. We have seen
that $m_{k+1}$ is the preferred son of~$m_k$ and that it is the penultimate son of that
latter node. So that from the root of $\cal W$, we have a branch whose nodes have
that additional property that the root excepted, all nodes are preferred sons of the 
previous node according to the son-father order. In fact many branches do possess a
similar property. Consider a node~$\nu$ and let ${\cal T}_{\nu}$ be the sub-tree
of $\cal W$ rooted at~$\nu$. We call {\bf \zzz-branch} the branch of~${\cal T}$
whose nodes, the root possibly excepted, have the signature \zzz, so that they
are the preferred son of the previous node. We can infer that property from the
recursive application of the rule applied to \www\zzz-nodes, namely the
rule \hbox{\www\zzz $\rightarrow$ \bbb\uu...\www\ccc\www\zzz\www\uu} which is applied
to the penultimate node in the right-hand side of the rule. The computations which
we performed in Lemma~\ref{lmwdecomp1} can be applied to the nodes 
$\rho_{\aa,n}$ for \hbox{\aa{} $\in\{\numd..\ddd,\uu\zzz\}$}. For such a node $\nu$,
which is of the form $M_n$$-$$km_n$, with $k\in\{0..p$$-$$4\}$, and whose
metallic code is \aa\zzz\uu$^{n-1}$. A node at the level~$h$ from~$\nu$
belonging to the \zzz-branch issued from~$\nu$, is at the distance $M_h$
from $\rho_{\aa,n+h}$. The computation gives the same result as the iterated 
application of the above rule: \aa\zzz\uu$^{n-1}$\zzz$^h$.

\section{Properties of the black metallic tree}\label{black_metal}

As defined in Subsection~\ref{smetalnum}, the black metallic tree $\cal B$ is defined
by the same rules as the white one, the difference being that the root of $\cal B$
is a black node. We know that the number of nodes on the level~$n$ of~$\cal B$ is
$b_n$ which satisfies $(3)$. We also know that \hbox{$B_n=m_n$}. Accordingly,
the nodes of the rightmost branch of $\cal B$ are numbered by $m_n$ and
their metallic code is \hbox{\uu\zzz$^n$}.

We can formulate an analogous version of Lemma~\ref{lmwdecomp1}.

\begin{lemm}\label{lmbdecomp1}
Let $\cal B$ be the black metallic tree dotted with its natural numbering. 
As for Lemma~{\rm\ref{lmwdecomp1}}, denote by ${\cal B}_{\numd,n}$, ${\cal W}_{\numt,n}$,
..., ${\cal W}_{\ddd,n}$ and ${\cal W}_{\uu\zzz,n}$ the metallic sub-trees of $\cal B$ of 
height~$n$ rooted at the nodes \numd, \numt, ..., \ddd{} and \uu\zzz{} the sons of the 
root \uu. Denote by $\varphi_{\numd,n}$, $\varphi_{\numt,n}$, ..., $\varphi_{\ddd,n}$ 
and $\varphi_{\uu\zzz,n}$ the rightmost node of the respective sub-trees 
on the level~$n$$+$$1$ of $\cal B$. We get that:
\vskip 5pt
\ligne{\hfill
$\vcenter{\vtop{\leftskip 0pt\hsize=220pt
	\ligne{{\bf [$\varphi_{\numd,n}$]} $=$ \uu\ccc$^{n-1}$\ddd, 
	{\bf [$\varphi_{\numt,n}$]} $=$ \numd\ccc$^{n-1}$\ddd,\hfill} 
	\ligne{\hfill ...,\hfill}
	\ligne{{\bf [$\varphi_{\ddd,n}$]} $=$ \ccc$^n$\ddd, 
	{\bf [$\varphi_{\uu\zzz,n}$]} $=$ \uu\zzz$^{n+1}$.\hfill} 
}}$
\hfill $(16)$\hskip 10pt}
\end{lemm}

\noindent
Proof. The proof is the same as for Lemma~\ref{lmwdecomp1}: we subtract $m_n$
from~$\varphi_{\uu\zzz,n}$ and we repeat until we reach $\varphi_{\numd,n}$. At each
step, we apply the subtraction using Algorithms~\ref{acomparemc}, \ref{acomplement},
\ref{aaddmetal} and \ref{alibre}.\hfill $\Box$

   We shall see that the properties of the sons signatures of the nodes in the black
metallic tree are different from those we have noted in the white one.
Figure~\ref{fmetalnoir} illustrates the black metallic tree for \hbox{$p=9$} as
in the case of Figure~\ref{fmetalblanc} to which the reader is referred for a comparison
between $\cal W$ and $\cal B$. We shall go back to that comparison in 
Section~\ref{scompnumwb}, illustrated by Figure~\ref{fmetalcomp} in that section.

\vskip 10pt
\vtop{
\ligne{\hfill
\includegraphics[scale=0.35]{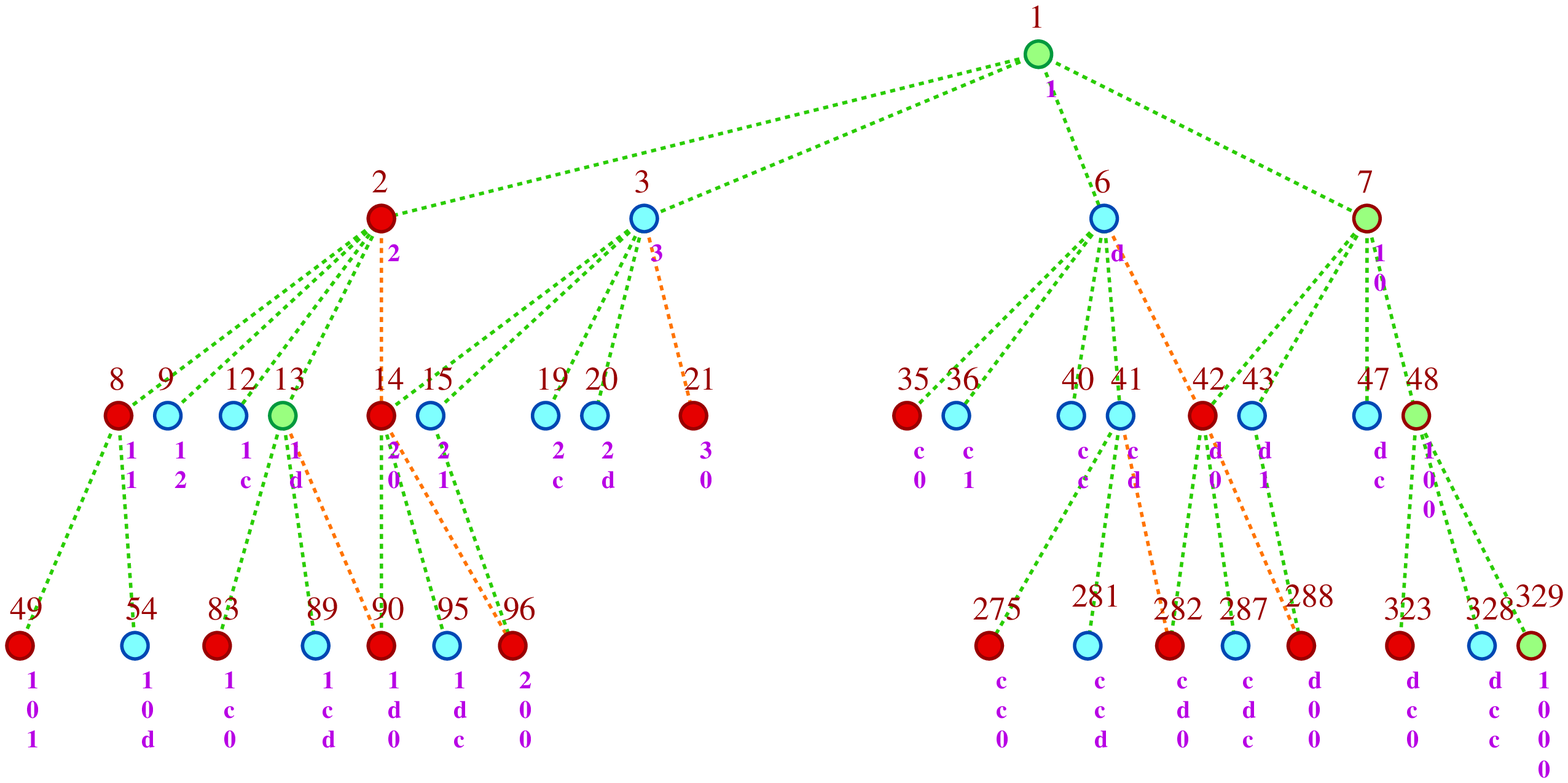}
\hfill}
\vspace{-10pt}
\ligne{\hfill
\vtop{\leftskip 0pt\parindent 0pt\hsize=300pt
\begin{fig}\label{fmetalnoir}
\leurre
The black metallic tree. The same convention about colours of the nodes and of the edges
between nodes as in Figure~{\rm\ref{fmetalblanc}} is used. 
We can see that the preferred son property as
stated in Theorem~{\rm\ref{tmpreferred}} is not true in the present setting.
\end{fig}
}
\hfill}
}
\vskip 5pt
Figure~\ref{fmetalnoir} shows us that the preferred is no more true. The leftmost son
of a level, a black node, has no son whose signature is~\zzz. All other nodes
have a son whose signature is~\zzz, and among them, the last node of a level has two
sons whose signature is~\zzz. Now, for a node~$\nu$ which has a unique son whose signature
is~\zzz, the metallic code of that node is not {\bf [$\nu$]0} but it is {\bf [$\mu$]0},
where \hbox{$\mu=\nu$$-$1}. Call {\bf successor} of the node~$\nu$, the node whose
metallic code is {\bf [$\nu$]0}. We can state:

\begin{thm}\label{tmblacksucc}
Define types for the nodes of a black metallic tree as follows:
\bbzz,\bbuu{} for a black node whose signature is~\zzz,\uu{} respectively,
\wwzz{} for a white node whose signature is~\zzz{}
and \wwaa{} for a white node whose signature is not~\zzz. We have
the following rules on the types of the nodes and the signatures~:
\vskip 5pt
\ligne{\hfill
$\vcenter{\vtop{\leftskip 0pt\hsize=300pt\bf
\ligne{\hfill \bbzz{} $\,\,\,\,\,\rightarrow$ \bbzz{}$($\wwaa$)^{p-4},$\hskip 20pt
\bbuu{} $\,\,\,\,\,\rightarrow$ \bbuu{}$($\wwaa$)^{p-4},$\hfill}
\ligne{\hfill \wwaa{} $\,\,\,\rightarrow$ \bbzz{}$($\wwaa$)^{p-3},$\hskip 20pt
\wwzz{} $\rightarrow$ \bbzz{}$($\wwaa$)^{p-4}$\wwzz$.$\hfill}
}}$
\hfill $(17)$\hskip 10pt}
\vskip 5pt
For any node~$\nu$ which is not the rightmost one on a level,
the successor of~$\nu$ is the leftmost node of~$\nu$$+$$1$. For the rightmost node
on the level~$n$, its successor is the rightmost node on the level~$n$$+$$1$.
We also have that the type~\bbuu{} occurs for the leftmost node of a level
only and that the type~\wwzz{} occurs for the rightmost node of a level only.
\end{thm}

\noindent
Proof. We again use Table~\ref{tsignatures}. But this time, the lines \fnb{1} and
\fnb{2} of the table will be used by all the nodes and the other lines of the table
will not be used.
\vskip 5pt
\vtop{
\begin{tab}\label{tmbsons}
Computation of the sons signatures in the black metallic tree. To left, $\nu$
which is supposed to observe the relations $(17)$. To right, the node $\nu$$+$$1$.
In the table, \hbox{\aa $\leq$ \ccc} if \gaa{} does not contain the suffix
{\bf \ddd\ccc$^\ast$} and the value \hbox{\aa=\ccc} is ruled out if
\gaa{} contains that suffix.
\end{tab}
\vskip-8pt
\ligne{\hskip 10pt \fnb 1\hskip 90pt \gaa\uu\hskip 110pt \gaa\numd\hfill}
\ligne{
\demiligne {\bbb\uu} {\gaa\zzz\uu} {...} {\gaa\zzz\ccc} {\gaa\zzz\ddd} {}
\demiligne {\www\numd} {\gaa\uu\zzz} {...} {\gaa\uu\ccc$^-$} {\gaa\uu\ccc} 
{\gaa\uu\ddd} 
\hfill}
\vskip 5pt
\ligne{\hskip 10pt \fnb 2\hskip 90pt \gaa\aa\hskip 110pt \gaa\aa$^+$\hfill}
\ligne{
\demiligne {\www\aa} {\gaa\aa$^-$\zzz} {...} {...} {\gaa\ccc$^-$\ccc} 
{\gaa\ccc$^-$\ddd}
\demiligne {\www\aa$^+$} {\gaa\aa\zzz} {...} {\gaa\aa\ccc$^-$} {\gaa\aa\ccc} 
{\gaa\aa\ddd} 
\hfill}
\vskip 5pt
\setbox150=\vtop{\leftskip 0pt\hsize=160pt
\ligne{\demiligne {\bbb\zzz} {\gaa\ddd\zzz} {...} {\gaa\ddd\ccc$^-$} {\gaa\ddd\ccc} 
{}\hfill}
\ligne{\demiligne {\www\zzz} {\gaa\ddd\zzz} {...} {\gaa\ddd\ccc$^-$} {\gaa\ddd\ccc} 
{\gaa$^+$\zzz\zzz}\hfill}
}
\setbox160=\vtop to \ht150{\leftskip 0pt\hsize=160pt
\vfill
\ligne{\demiligne {\www\ddd} {\gaa\ccc\zzz} {...} {...} {\gaa\ccc\ccc} 
{\gaa\ccc\ddd}\hfill}
\vfill
}
\ligne{\hskip 10pt \fnb 3\hskip 90pt \gaa\ddd\hskip 110pt \gaa$^+$\zzz\hfill}
\ligne{\box160\hfill\box150 \hfill}
\vskip 5pt
\setbox150=\vtop{\leftskip 0pt\hsize=160pt
\ligne{\demiligne {\bbb\zzz} {\gaa\ccc\zzz} {...} {\gaa\ccc\ccc$^-$} 
{\gaa\ccc\ccc} {}\hfill}
\ligne{\demiligne {\www\zzz} {\gaa\ccc\zzz} {...} {\gaa\ccc\ccc$^-$} {\gaa\ccc\ccc} 
{\gaa$^+$\zzz\zzz}\hfill}
}
\setbox160=\vtop to \ht150{\leftskip 0pt\hsize=160pt
\vfill
\ligne{\demiligne {\www\ccc} {\gaa\ccc$^-$\zzz} {...} {...} {\gaa\ccc$^-$\ccc} 
{\gaa\ccc$^-$\ddd}\hfill}
\vfill
}
\ligne{\hskip 10pt \fnb 4\hskip 90pt \gaa\ccc\hskip 110pt \gaa$^+$\zzz\hfill}
\ligne{\box160\hfill\box150 \hfill}
\vskip 5pt
\setbox150=\vtop{\leftskip 0pt\hsize=160pt
\ligne{\demiligne {\bbb\zzz} {\gaa$^-$\ccc\zzz} {...} {...} 
{\gaa$^-$\ccc\ccc} {}\hfill}
\ligne{\demiligne {\bbb\zzz} {\gaa$^-$\ddd\zzz} {...} {...} 
{\gaa$^-$\ddd\ccc} {}\hfill}
\ligne{\demiligne {\www\zzz} {\gaa$^-$\ccc\zzz} {...} {\...} {\gaa$^-$\ccc\ccc} 
{\gaa\zzz\zzz}\hfill}
\ligne{\demiligne {\www\zzz} {\gaa$^-$\ddd\zzz} {...} {\...} {\gaa$^-$\ddd\ccc} 
{\gaa\zzz\zzz}\hfill}
}
\setbox160=\vtop to \ht150{\leftskip 0pt\hsize=160pt
\vskip -2pt
\ligne{\demiligne {\www\uu} {\gaa\zzz\zzz} {...} {\gaa\zzz\ccc$^-$} {\gaa\zzz\ccc} 
{\gaa\zzz\ddd}\hfill}
\vskip 12pt
\ligne{\demiligne {\bbb\uu} {\gaa\zzz\uu} {...} {\gaa\zzz\ccc} {\gaa\zzz\ddd} 
{}\hfill}
\vfill
}
\ligne{\hskip 10pt \fnb 5\hskip 90pt \gaa\zzz\hskip 110pt \gaa\uu\hfill}
\ligne{\box150\hfill\box160 \hfill}
%}
%\hfill}
}
\vskip 10pt
We can see that under the assumptions of~$(17)$ applied to the left-hand side of the
table, the right hand-side also observes the rules of~$(17)$. 
Table~\ref{tmbsons} also shows that the position of the successor is that which is
indicated in the statement of the theorem. We can also see that what is said in that
statement for the \bbb\uu-nodes and for the \www\zzz-ones is observed.
Accordingly, Theorem~\ref{tmblacksucc} is proved.\hfill $\Box$

Note that we can also say for the black metallic tree that $m_{k+1}$ is the
preferred son of~$m_k$.

\subsection{Connection of the white metallic tree with the tilings $\{p,4\}$
and $\{p$+$2,3\}$ of the hyperbolic plane}~\label{stilings}

   As mentioned in the introduction, the white metallic tree is connected with the
tilings $\{p,4\}$ and $\{p$+$2,3\}$ of the hyperbolic plane with \hbox{$p\geq5$}. 
Those tilings are generated by the reflection of
a basic polygon in its sides and the recursive reflections of the images in their sides.
The basic polygon is the regular convex polygon with $p$, $p$+2~sides and 
with $\displaystyle{\pi\over2}$,$\displaystyle{{2\pi}\over3}$ as vertex angle
in $\{p,4\}$,$\{p$+$2,3\}$ respectively. 

\vskip 10pt
\vtop{
\ligne{\hfill
\includegraphics[scale=0.6]{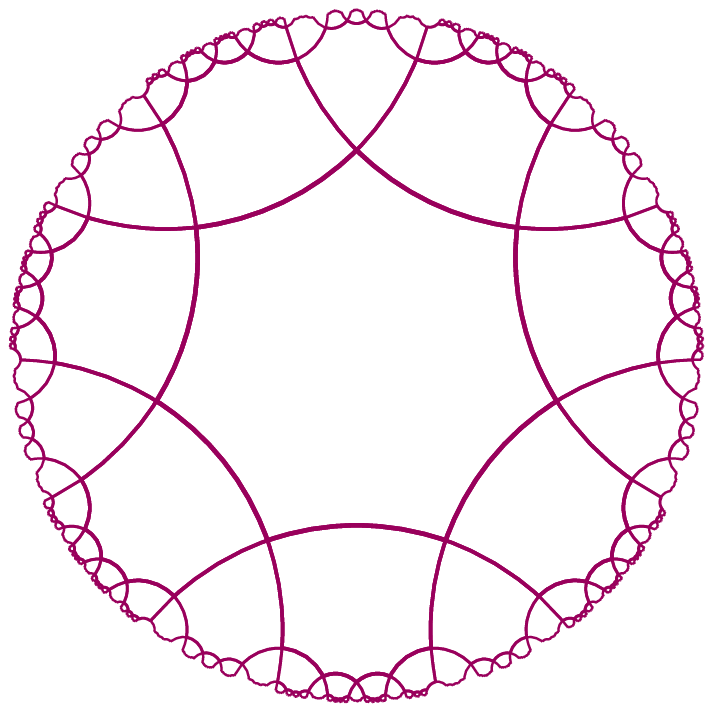}
\includegraphics[scale=0.6]{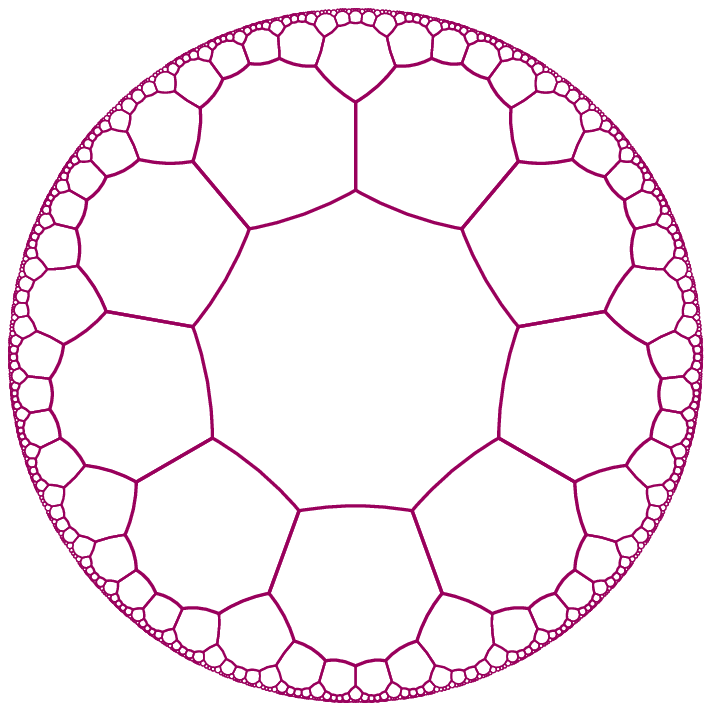}
\hfill}
\vspace{-10pt}
\ligne{\hfill
\vtop{\leftskip 0pt\parindent 0pt\hsize=300pt
\begin{fig}\label{f7_9til}
\leurre
The tilings generated by the white metallic tree with $p=7$.
To left, the the tiling $\{7,4\}$ to right, the tiling $\{9,3\}$
\end{fig}
}
\hfill}
}
\vskip 10pt
\vtop{
\ligne{\hfill
\includegraphics[scale=0.6]{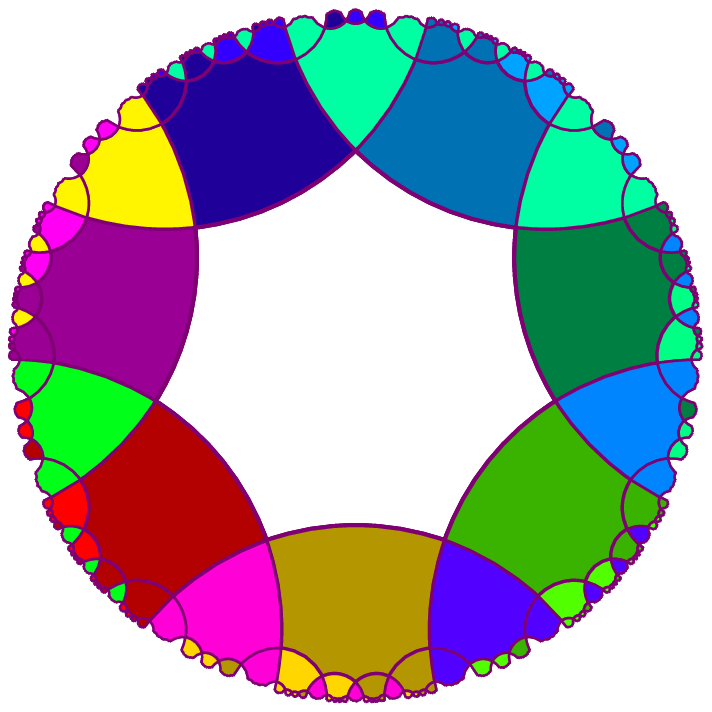}
\includegraphics[scale=0.6]{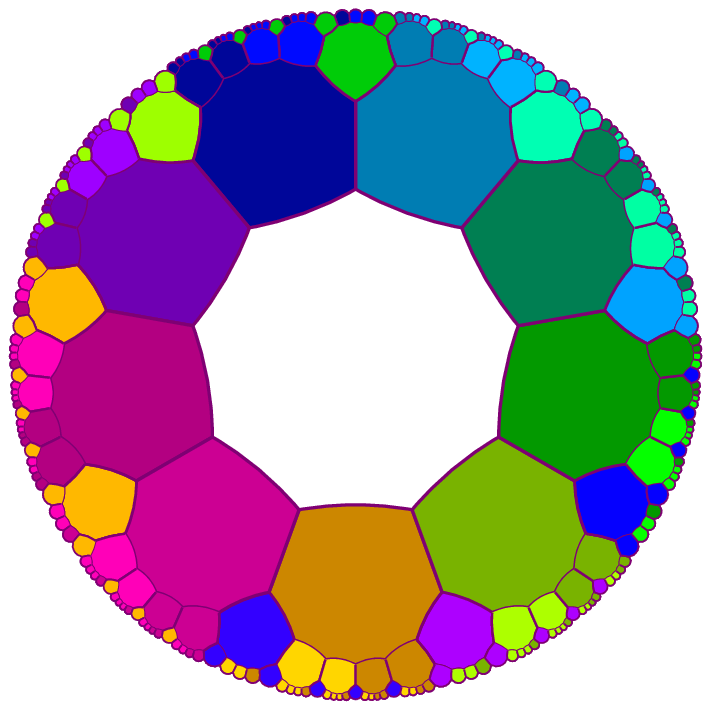}
\hfill}
\vspace{-10pt}
\ligne{\hfill
\vtop{\leftskip 0pt\parindent 0pt\hsize=300pt
\begin{fig}\label{fspan7_9til}
\leurre
How the white metallic tree generates the tilings $\{7,4\}$ and $\{9,3\}$:
the sectors are delimited by colours, each sector being associated with three colours
which are attached to the status of the nodes. Each sector in the above figures is 
spanned by the white metallic tree.
\end{fig}
}
\hfill}
}
\vskip 5pt
\vskip 5pt
Those polygons
live in the hyperbolic plane, not in the Euclidean one. 
Figure~\ref{f7_9til} illustrates the tiling $\{7,4\}$, left hand side, and the
tiling $\{9,3\}$, right hand side, associated to $p=7$.

Figure~\ref{fspan7_9til} illustrates how the white metallic tree generates
the considered tilings. In both tilings, the tiles of a {\bf sector}, can be put in
bijection with the nodes of the tree. In the case of the tiling $\{p,4\}$, such a 
sector is a quarter of the plane: it is delimited by two perpendicular half-lines 
stemming from
the same vertex~$V$ of a tile~$\tau$ and passing through the other ends of the edges
of~$\tau$ sharing~$V$.
The definition of the sector is more delicate in the case of the tiling 
$\{p$+$2,3\}$. 
The sector
is also defined by half-lines which are, this time, issued from the mid-point of an 
edge~$\eta$ and those half-lines pass through the mid-points of two consecutive sides of 
a tile sharing~$V$ as a vertex, $V$ being also an end of~$\eta$. The reader is
referred to~\cite{mmbook1} for proofs of the just mentioned properties. Accordingly,
as shown on the figure illustrating the case when $p=7$, seven sectors allow us to 
locate tiles in the tiling $\{p,4\}$ and
nine sectors allow us to perform the same thing in the tiling $\{p$+$2,3\}$. From now on,
we call {\bf tile $\nu$} the tile attached to the node~$\nu$ of such a white metallic 
tree we assume to be fixed once and for all. We also say that $[\nu]$ is the code 
of the tile~$\nu$.

In Sub-section~\ref{smwneighpathpp23}, we shall prove that the
preferred son property allows us to compute in linear time with respect to the code
of a node~$\nu$ the codes of the nodes attached to the tiles which share a side 
with the tile~$\nu$. Such tiles are called the {\bf neighbours} of~$\nu$. 
We shall also see that Theorem~\ref{tmpreferred} allows us to compute in linear time 
with respect to
$[\nu]$ a shortest path in the tiling, leading from the tile~$\nu$ to tile~1.

\subsection{The neighbours of a node in $\{p,4\}$}
\label{smwneighpathp4}

   Consider a tile~$\tau$ in the tiling $\{p,4\}$. Fix a central tile which will be 
numbered by~0 and fix $p$~sectors around tile~0. Another tile is a {\bf neighbour}
of~$\tau$ if and only of it shares a side of~$\tau$. In order to identify the neighbours
of~$\tau$, we number its sides as follows: side~1 is the side shared with the father
of~$\tau$ in the white metallic tree which spans the sector to which $\tau$ belongs.
By definition, the father of the leading tile of a sector is tile~0. Tile~0 has no father
but we number its side by fixing its side~1 once and for all. Now that the side~1 of 
each tile
is defined, we number the other sides while counterclockwise turning around the tile,
giving the number $n$+1 when meeting the new side after the side~$n$. Accordingly,
the $p$ neighbours of~$\tau$ are numbered from~1 up to~$p$. The tile which shares with 
$\tau$ its side~$i$ is called the {\bf neighbour}~$i$ of $\tau$ and we denote it by 
$\tau_i$. Accordingly, $\tau_1$ is the father of~$\tau$, except when \hbox{$\tau = 0$}.
Thanks to Theorem~\ref{tmpreferred}, we indicate how to compute the metallic code
of $\tau_i$ for each~$i$. It will also help us to construct the path from~$\tau$
to tile~0.

\def\til#1{\hbox{$\boxright$#1$\boxleft$}}
We shall identify a tile with its number in its sector and also by the metallic code of
its number. If $\tau$ is the tile, $n(\tau)$ is its number, {\bf [$\tau$]} is its
metallic code and \til{$n$}{} is the tile whose number is~$n$. If $\tau$ is a white 
node numbered by~$\nu$, its sons are $\tau_i$ 
with \hbox{$i\in[2..p$$-$$1]$}. We can see in~Figure~\ref{fmetalblanc} that $\tau_p$ is 
\setbox110=\hbox{\til{$n(\tau)$+1}}
the leftmost son of~$\nu$+1. We can write: \hbox{$\tau_p =$ ${\copy110}_{\alpha}$},
with $\alpha=2,3$ depending on whether $\nu$+1 is white, black respectively.
\setbox120=\hbox{\til{$n(\tau)$$-$1}}
If $\tau$ is a black node, \hbox{$\tau_2 = \box120$},
its sons are $\tau_i$ for \hbox{$i\in[3..p$$-$$1]$} and $\tau_p$ is again
the leftmost son of~$\nu$+1, so that we have \hbox{$\tau_p =$ ${\copy110}_2$}
as $\nu$+1 is a white node.

We turn now to the writing of the metallic codes for the neighbours of a tile~$\tau$
from {\bf [$\tau$]}. Note that if \hbox{\bf [$\tau$] $=$ a$_k$$..$a$_0$}
and \hbox{\bf [$\tau$$-$$1$] $=$ b$_k$$..$b$_0$}, we have:

\begin{lemm}\label{lmwneighp4}
Let $\tau$ be a node and let \hbox{\bf [$\tau$] $=$ a$_k$$..$a$_0$}.
Let \hbox{\bf [$\tau$]$\ominus$\uu{} $=$ b$_k$$..$b$_0$}. Then,
\vskip 5pt
\ligne{\hfill
$\vcenter{\vtop{\leftskip 0pt\hsize=205pt\bf
\ligne{\hfill \rm if $\tau$ is a \wwaa-node:\hfill}
\vskip 2pt
\ligne{\hfill[$\tau_1$] $=$ b$_k$$..$b$_1$,\hskip 20pt
	$[\tau_i]$ $=$ b$_k$$..$b$_0$[$i$$-$$1$]$,\ i\in[2..p$$-$$2],$\hfill}
	\ligne{\hfill[$\tau_{p-1}$] $=$ a$_k$$..$a$_0$\zzz,\hskip 20pt 
	$[\tau_p]$ $=$ a$_k$$..$a$_0$\uu.\hfill}
\vskip 2pt
\ligne{\hfill \rm if $\tau$ is a \bbb-node:\hfill}
\vskip 2pt
\ligne{\hfill[$\tau_1$] $=$ b$_k$$..$b$_1$,\hskip 20pt
       [$\tau_2$] $=$ b$_k$$..$b$_1$$\ominus$\uu,\hfill}
	\ligne{\hfill$[\tau_i]$ $=$ b$_k$$..$b$_0$[$i$$-$$1$]$,\ i\in[3..p$$-$$2],$\hfill}
	\ligne{\hfill[$\tau_{p-1}$] $=$ a$_k$$..$a$_0$\zzz,\hskip 20pt 
	[$\tau_p$] $=$ a$_k$$..$a$_0$\uu.\hfill}
\vskip 2pt
\ligne{\hfill \rm if $\tau$ is a \www$\alpha$-node, $\alpha\in[\zzz,\uu]$:\hfill}
\vskip 2pt
        \ligne{\hfill[$\tau_1$] $=$ a$_k$$..$a$_1$$,$\hskip 20pt
	%\ligne{\hfill
	[$\tau_i$] $=$ b$_k$$..$b$_0$[$i$$-$$1$+$\alpha$]$,
                     \ i\in[2..p$$-$$3]$$,$\hfill}
	\ligne{\hfill[$\tau_{p-2}$] $=$ a$_k$$..$a$_0$\zzz,\hskip 20pt 
	[$\tau_{p-1}$] $=$ a$_k$$..$a$_0$\uu,\hfill}
	\ligne{\hfill[$\tau_p$] $=$ a$_k$$..$a$_0$\uu.\hfill}
}}
$
\hfill$(18)$\hskip 10pt}
\end{lemm}

\noindent
Proof: the proof is a direct application of the relation~$(14)$.\hfill $\Box$

We can conclude from~$(18)$ the following algorithm to compute {\bf [$\tau_1$]}
from~{\bf [$\tau$]}:

\vskip-5pt
\def\oor{{\bf or}}
\vtop{
\begin{algo}\label{aperep4}
Computation of the father of~$\tau$ from {\bf [$\tau$]}.
We assume that \hbox{\bf [$\tau$] $=$ a$_k$$..$a$_0$} and
	that $(\tau)_i = \aa_i$, with \hbox{$i\in\{0..k\}$}.
\end{algo}
\vskip-8pt
\ligne{\hfill
\vtop{\leftskip 0pt\hsize=200pt
\hrule height 0.3pt depth 0.3pt width \hsize
\vskip 5pt
\ligne{\hskip 10pt \iff{} $(\tau)_0 \in \{\numd..\ddd\}$\hfill}
\ligne{\hskip 10pt \hskip 10pt \tthen{}
{\bf [$\tau_1$]} := {\bf a$_k$$..$a$_1$$\oplus$\uu};\hfill}
\ligne{\hskip 10pt \hskip 10pt \eelse{} \iff{} $(\tau)_0$ = \zzz\hfill}
\ligne{\hskip 10pt \hskip 10pt \hskip 30pt \tthen{}
{\bf [$\tau_1$]} := {\bf a$_k$$..$a$_1$};\hfill}
\ligne{\hskip 10pt \hskip 10pt \hskip 30pt \eelse{}
$i$ := 1; \hfill}
\ligne{\hskip 10pt \hskip 10pt \hskip 30pt \hskip 20pt\wwhile{}
$(\tau)_i$ = \uu{} \oor{} $i$ $>$ $k$\hfill}
\ligne{\hskip 10pt \hskip 10pt \hskip 30pt \hskip 20pt\hskip 5pt
\lloop{} $i$ := $i$+1; \endloop;\hfill}
\ligne{\hskip 10pt \hskip 10pt \hskip 30pt \hskip 20pt\iff{} $i$ $>$ $k$ \oor\eelse{}
$(\tau)_i$ $=$ \zzz\hfill}
\ligne{\hskip 10pt \hskip 10pt \hskip 30pt \hskip 20pt \hskip 10pt \tthen{}
{\bf [$\tau_1$]} := {\bf a$_k$$..$a$_1$};\hfill}
\ligne{\hskip 10pt \hskip 10pt \hskip 30pt \hskip 20pt \hskip 10pt \eelse{}
{\bf [$\tau_1$]} := {\bf a$_k$$..$a$_1$$\oplus$\uu};\hfill}
\ligne{\hskip 10pt \hskip 10pt \hskip 30pt \hskip 20pt\endif;\hfill}
\ligne{\hskip 10pt \hskip 10pt \hskip 20pt \endif;\hfill}
\ligne{\hskip 10pt \endif;\hfill}
\vskip 5pt
\hrule height 0.3pt depth 0.3pt width \hsize
}
\hfill}
}
\vskip 5pt
\vskip 5pt
The first condition in the algorithm comes from the examination of the rules giving the
sons signatures. It is clear when \hbox{\sgn{}$(\tau) \in \{\numt..\ddd\}$}. When
\hbox{\sgn{}$(\tau) = \numd$}, whether the node is black or white, it is the leftmost
son of $\tau_1$ or its second son respectively. And so, as the successor of~$\tau_1$
is its rightmost son, we have to perform what the algorithm indicates. When
\hbox{\sgn{}$(\tau) =\zzz$}, it is clear that {\bf [$\tau_1$]} is obtained as indicated
in the algorithm. We remain with the case when \hbox{\sgn{}$(\tau) = \uu$}. It is 
either a black node, in which case the father is given by 
\hbox{\bf a$_k$$..$a$_1$$\oplus$\uu}, or it is white but in that case the father
is \hbox{\bf a$_k$$..$a$_1$}. The difference of the situation is defined by the digits
which is to the left of {\bf a$_0$}. As long as we meet \uu{} while going to the left,
we cannot distinguish between the two cases. When we meet \aa$_i$ with 
\hbox{\aa$_i \not=$ \zzz}, we know that we are in the case of a black node.
If \hbox{\aa$_i =$ \zzz}, we are in the case of a white node: it is a corollary
of what was proved in Lemma~\ref{lmwdecomp1} and of the remarks we made after the
proof of Theorem~\ref{tmpreferred}. This completes the proof of the algorithm.
\hfill $\Box$

Algorithm~\ref{aperep4} is the key for devising an algorithm to compute a path
from a tile~$\tau$ to the leading tile of the sector where it lies. We cannot use the 
function defined by the algorithm as is. If we do that, in case the
metallic code contains a large pattern \uu$^\ast$, we have to repeat the \wwhile{}
\lloop{} each time we meet \uu{} which leads to a quadratic time.
The idea is to fix the choice of the definition of the father once \uu{} is detected.
Once we find the non \uu-digit of highest rank with respect to those \uu-digits,
we fix the choice accordingly until the pattern \uu$^\ast$ is dealt with.

Here is the algorithm:

%\vskip 10pt
\vtop{
\begin{algo}\label{apathp4}
Computation of the sequence of tiles which constitutes the path, along a branch of the
$\cal W$ from a given tile~$\tau$ to the leading tile of the sector which 
contains~$\tau$. We assume that \hbox{\bf [$\tau$] $=$ a$_k$$..$a$_0$}.
\end{algo}
\vskip-8pt
\ligne{\hfill
\vtop{\leftskip 0pt\hsize=300pt
\hrule height 0.3pt depth 0.3pt width \hsize
\vskip 5pt
\ligne{\hskip 10pt list := {\bf [$\tau$]}; r := 0; fixed := \faux; 
node := {\bf a$_k$$..$a$_0$};\hfill}
\ligne{\hskip 10pt \ffor{} $i$ \iin [0..$k$]\hfill}
\ligne{\hskip 10pt \hskip 10pt \lloop{}\hfill}
\ligne{\hskip 10pt \hskip 10pt \hskip 20pt \iff{} (node)$(i) \in \{\numd..\ddd\}$\hfill}
\ligne{\hskip 10pt \hskip 10pt \hskip 10pt \hskip 20pt \tthen{}
node := {\bf a$_k$$..$a$_{i+1}$$\oplus$\uu};\hfill}
\ligne{\hskip 10pt \hskip 10pt \hskip 10pt \hskip 20pt
\eelse{} \iff{} (node)$(i)$ = \zzz\hfill}
\ligne{\hskip 10pt \hskip 10pt \hskip 10pt \hskip 20pt \hskip 30pt \tthen{}
node := {\bf a$_k$$..$a$_{i+1}$};\hfill}
\ligne{\hskip 10pt \hskip 10pt \hskip 10pt \hskip 20pt \hskip 30pt \eelse{}
$i$ := 1; \hfill}
\ligne{\hskip 10pt \hskip 10pt \hskip 10pt \hskip 20pt \hskip 30pt \hskip 20pt 
\iff{} \nnot{} fixed\hfill}
\ligne{\hskip 10pt \hskip 10pt \hskip 10pt \hskip 20pt \hskip 30pt \hskip 20pt
\hskip 10pt \tthen{} r := $i$;\hfill}
\ligne{\hskip 10pt \hskip 10pt \hskip 10pt \hskip 20pt \hskip 30pt \hskip 20pt
\hskip 10pt \hskip 20pt\wwhile{} (node)$({\rm r})$ $=$ \uu{} \oor{} r $>$ $k$\hfill}
\ligne{\hskip 10pt \hskip 10pt \hskip 10pt \hskip 20pt \hskip 30pt \hskip 20pt
\hskip 10pt \hskip 20pt \hskip 10pt \lloop{} r := r+1; \endloop;\hfill}
\ligne{\hskip 10pt \hskip 10pt \hskip 10pt \hskip 20pt \hskip 30pt \hskip 20pt
\hskip 10pt \hskip 20pt \iff{} r $>$ $k$ \oor\eelse{} (node)$({\rm r})$ $=$ \zzz\hfill}
\ligne{\hskip 10pt \hskip 10pt \hskip 10pt \hskip 20pt \hskip 30pt \hskip 20pt
\hskip 10pt \hskip 20pt \hskip 10pt \tthen{}
node := {\bf a$_k$$..$a$_{i+1}$}; fixed := \vrai;\hfill}
\ligne{\hskip 10pt \hskip 10pt \hskip 10pt \hskip 20pt \hskip 30pt \hskip 20pt
\hskip 10pt \hskip 20pt \hskip 10pt \eelse{}
node := {\bf a$_k$$..$a$_{i+1}$$\oplus$\uu}; fixed := \faux;\hfill}
\ligne{\hskip 10pt \hskip 10pt \hskip 10pt \hskip 20pt \hskip 30pt \hskip 20pt
\hskip 10pt \hskip 20pt\endif;\hfill}
\ligne{\hskip 10pt \hskip 10pt \hskip 10pt \hskip 20pt \hskip 30pt \hskip 20pt
\hskip 10pt \eelse{} node := {\bf a$_k$$..$a$_{i+1}$};
\hfill}
\ligne{\hskip 10pt \hskip 10pt \hskip 10pt \hskip 20pt \hskip 30pt \hskip 20pt
\endif;\hfill}
\ligne{\hskip 10pt \hskip 10pt \hskip 10pt \hskip 20pt \hskip 20pt \endif;\hfill}
\ligne{\hskip 10pt \hskip 10pt \hskip 20pt \endif;\hfill}
\ligne{\hskip 10pt \hskip 10pt \hskip 20pt list := node \& list;\hfill}
\ligne{\hskip 10pt \endloop;\hfill}
\vskip 5pt
\hrule height 0.3pt depth 0.3pt width \hsize
}
\hfill}
}
\vskip 10pt
Note that this algorithm allows us to compute the path as a sequence of nodes.
The computation of each node requires at most $k$+1 digits, where $k$+1 is the initial
length of {\bf [$\tau$]}. The \ffor{}-loop has $k$+1 steps, so that the time complexity
of the computation is quadratic: it is $\alpha.(k$+$1)^2$. The space complexity is
also $\alpha.(k$+$1)^2$: the length of a digit is constant and we have also to take into
account a separator.

\subsection{The neighbours of a node in $\{p$+$2,3\}$}
\label{smwneighpathpp23}

After Subsection~\ref{smwneighpathp4}, we deal with the same question in the tiling
\hbox{$\{p$+$2,3\}$}. Note that there is no change for what concerns the spanning tree.
We already mentioned in Subsection~\ref{stilings} that the white metallic tree
spans the sectors in both tilings~$\{p,4\}$ and $\{p$+$2,3\}$. The difference, from
the point of view of the trees lies in the number of sectors: $p$ in the tiling
$\{p,4\}$, $p$+2{} in the tiling $\{p$+$2,3\}$.

That difference comes from a deeper one: the number of tiles around a vertex: 
4 of them in $\{p,4\}$ while there are 3~of them in $\{p$+$2,3\}$. That
difference in the number of tiles around a vertex has a consequence on the 
neighbourhood of a tile. We keep the same definition as for the tiling $\{p,4\}$:
a neighbour of a tile~$\tau$ is a tile which shares a side with~$\tau$. Of course,
a neighbour of~$\tau$ also shares two vertices with~$\tau$: the ends of the common
side. Now, in the tiling $\{p$+$2,3\}$, if two tiles share a vertex they also
share a side. It is also the reason while the tree for $\{p,4\}$ is the same
as the tree for $\{p$+$2,3\}$ and not for $\{p,3\}$ which is completely different.
So, in $\{p$+$2,3\}$, a tile has $p$+2 neighbours. Outside the father, the sons
and the son of another tile, $\tau$ has also two additional neighbours which lie
on the same level of the tree as $\tau$. We adopt the same numbering of the sides
of a tile as in Subsection~\ref{smwneighpathp4}. Side~1 being the side of the
father and $\tau$ being identified by its number, $\tau_2$ is \til{$\tau$$-$1},
the sons are \til{$\tau_i$} for \hbox{$i\in\{3..p\}$} for a white node,
for \hbox{$i\in\{4..p\}$} for a black one, $\tau_{p+1}$ is the leftmost son
of \til{$\tau$+1} and $\tau_{p+2}$ is \til{$\tau$+1}, the other tile which is on the
same level of the tree as~$\tau$.

This leads us to a modified version of Lemma~\ref{lmwneighp4}:

\def\ff{{\bf f}}
\begin{lemm}\label{lmwneighpp23}
Let $\tau$ be a node and let \hbox{\bf [$\tau$] $=$ a$_k$$..$a$_0$}.
Let \hbox{\bf [$\tau$]$\ominus$\uu{} $=$ b$_k$$..$b$_0$}. Then,
\vskip 5pt
\ligne{\hfill
$\vcenter{\vtop{\leftskip 0pt\hsize=250pt\bf
\ligne{\hfill \rm if $\tau$ is a \wwaa-node:\hfill}
\vskip 2pt
	\ligne{\hfill[$\tau_1$] $=$ b$_k$$..$b$_1$$,$ \hskip 10pt
                     [$\tau_2$] = [$\tau$] $\ominus$ \uu$,$\hskip 10pt
	[$\tau_i$] $=$ b$_k$$..$b$_0$[$i$$-$$2$]$,\ i\in[3..p$$-$$1],$\hfill}
	\ligne{\hfill[$\tau_p$] $=$ a$_k$$..$a$_0$\zzz$,$\hskip 10pt 
	[$\tau_{p+1}$] $=$ a$_k$$..$a$_0$\uu$,$\hskip 10pt 
	[$\tau_{p+2}$] $=$ [$\tau$] $\oplus$ \uu$.$\hfill}
\vskip 2pt
	\ligne{\hfill \rm if $\tau$ is a \bbb-node:\hfill}
\vskip 2pt
\ligne{\hfill[$\tau_1$] $=$ b$_k$$..$b$_1$$,$\hskip 10pt
	[$\tau_2$] $=$ [$\tau_1$] $\ominus$ \uu$,$\hskip 10pt
	[$\tau_3$] $=$ [$\tau$] $\ominus$ \uu$,$\hskip 10pt
       \hfill}
	\ligne{\hfill$[\tau_i]$ $=$ b$_k$$..$b$_0$[$i$$-$$2$]$,\ i\in[4..p$$-$$1],$\hfill}
	\ligne{\hfill[$\tau_p$] $=$ a$_k$$..$a$_0$\zzz,\hskip 10pt 
	[$\tau_{p+1}$] $=$ a$_k$$..$a$_0$\uu$,$\hskip 10pt
	[$\tau_{p+2}$] $=$ [$\tau$] $\oplus$ \uu$.$\hfill}
\vskip 2pt
\ligne{\hfill \rm if $\tau$ is a \www$\alpha$-node, $\alpha\in[\zzz,\uu]$:\hfill}
\vskip 2pt
        \ligne{\hfill[$\tau_1$] $=$ a$_k$$..$a$_1$$,$\hskip 10pt
		     [$\tau_2$] $=$ [$\tau$] $\ominus$ \uu$,$\hskip 10pt
	[$\tau_i$] $=$ b$_k$$..$b$_0$[$i$$-$$2$+$\alpha$]$,
                     \ i\in[3..p$$-$$2$]$,$\hfill}
	\ligne{\hfill[$\tau_{p-1}$] $=$ a$_k$$..$a$_0$\zzz$,$\hskip 10pt 
	[$\tau_p$] $=$ a$_k$$..$a$_0$\uu$,$\hfill}
	\ligne{\hfill[$\tau_{p+1}$] $=$ a$_k$$..$a$_0$\numd$.$\hskip 10pt
	[$\tau_{p+2}$] $=$ [$\tau$] $\oplus$ \uu$,$\hfill}
}}
$
\hfill$(19)$\hskip 10pt}
\end{lemm}

The lemma is not different from Lemma~\ref{lmwneighp4}. It is the reason why
we may apply Algorithm~\ref{apathp4} to the tilings $\{p$+$a,3\}$ without any change.
However, we indicate here another algorithm which also work for the tilings $\{p,4\}$.
Note that Algorithm~\ref{apathp4} is a bottom-up algorithm: it constructs the
path from the tile to the leading one by scrutinizing the digits of the metallic
code from its weakest metallic component up to its highest one. We constructed the
path as a LIFO-stack. Here, we proceed in the reverse order: from the highest component 
down to the weakest one. Accordingly, the path will be constructed as a FIFO-stack.
The first tile of the path is, of course, the leading one. Next, we look at the
highest digit: in most cases, it indicates in which sub-tree ${\cal B}_{\numd}$
or ${\cal W}_{\aa}$ the tile belongs, where \hbox{\aa{} $\in \{$\numt$..$\zzz\uu $\}$}.
But in some cases, we might hesitate between ${\cal W}_{\aa}$ 
and ${\cal W}_{\aa\oplus\uu}$ or between ${\cal W}_{\uu\uu}$ and ${\cal B}_{\numd}$.
As an example, {\bf \uu$^k$\numd} belongs to ${\cal B}_{\numd}$ while
{\bf \uu$^k$} belongs to ${\cal W}_{\uu\uu}$. That example indicates that the
ambiguity can be raised by reading the last digit only. In order to raise
the ambiguity, we chose both sub-trees and we repeat that kind of choice at each step.
Now, let us show that no binary tree is raised in that process. Assume that we have
a single path \hbox{$\pi = \{\pi_0..\pi_k\}$} and that reading the digit~\aa, we 
hesitate between the sub-trees $T_{\mu}$ and $T_{\mu\oplus\uu}$. We constitute
two paths: \hbox{$\pi = \{pi_0..\pi_k\mu\}$} and \hbox{$\omega = \{\pi_0..\pi_k\mu^+\}$},
where we denote $\mu\oplus\uu$ by $\mu^+$. We have the condition 
\hbox{$0\leq\mu^+-\mu\leq1$}. Now let \bbb{} be the next digit we read.
Assume that starting from $\mu$, we hesitate between two sub-trees rooted at two
consecutive sons of~$\mu$, $\nu$ and $\nu^+=\nu\oplus1$.
Similarly, \bbb{} gives raise to the possible choices between consecutive sons
of $\mu^+$, say $\varphi$ and $\varphi^+=\varphi\oplus1$. Now, 
we can see that \hbox{$(\aa\bbb)<(\aa^+\zzz)$}. It means that the tile cannot be both 
in the
sub-tree rooted at~$\nu^+$ and that rooted at~$\varphi^+$: in between them there is
the tree rooted at $\varphi$. A similar argument tells us that the tile cannot
be both in the sub-tree rooted at~$\nu$ and that rooted at~$\varphi$, the sub-tree 
rooted at $\nu^+$ lying in between them. So we may continue either by appending
$\nu$ and $\nu^+$ to $\pi$, or by appending $\varphi$ and $\varphi^+$ to $\omega$
or by appending $\nu^+$ to~$\pi$ and $\varphi$ to~$\omega$. In all cases
the distance between two nodes belonging to each path with the same rank being~1,
except at the initialisation step and at the end of the algorithm.
From that we get Algorithm~\ref{apathpp23}.

   In order to better understand the algorithm, we indicate several of its features.
The metallic code of the node for which we compute the path to the root is represented as
a table whose elements are the digits of the code. The path is represented by a table 
whose elements are digits and a letter, \www{} or \bbb, indicating the status of the
node. We start from the root which does not occur in the table. Recursively, the status
indicated at the considered entry helps us to know which son is represented in the next
case which indicates the signature of that latter node. This is a difference with 
Algorithm~\ref{apathp4}. In the bottom-up approach, we do not know the status of the
current node~$\nu$. If the signature of~$\nu$ is in \hbox{\zzz,\numt..\ddd}, we know its
status, otherwise it is not possible without further information. 

Let us look at what can be given by the top-down approach.
Let ~$\mu$ be a node. The signature of a son~$\nu$ of~$\mu$ and
the status of~$\mu$ allows us to identify~$\nu$ and to know its status, so that
we can recursively continue the identification of the nodes on the path using
the successive digits of~{\bf [$\nu$]}. Let us look at the way to do that precisely.
We start from the root, and we compute two tables list$^-$
\ligne{\hfill}
\vskip 10pt
\def\ccase{{\bf case}}
\def\iis{{\bf is}}
\def\wwhen{{\bf when}}
\def\endcase{{\bf end case}}
\def\eet{{\bf and}}
\vtop{
\begin{algo}\label{apathpp23}
Computation of the sequence of tiles which constitutes the path, along a branch of 
$\cal W$ from a given tile~$\tau$ to the leading tile of the sector which 
contains~$\tau$. We assume that \hbox{\bf [$\tau$] $=$ a$_k$$..$a$_0$}. The path
is represented by a table whose elements are a number, the signature of the considered
node, together with its status. 
\end{algo}
\vskip-8pt
\ligne{\hfill
\vtop{\leftskip 0pt\hsize=330pt
\hrule height 0.3pt depth 0.3pt width \hsize
\vskip 5pt
\ligne{\hskip 10pt node := {\bf a$_k$$..$a$_0$};\hfill}
\ligne{\hskip 10pt list$^-$(1) := \aa$_k$\www; 
list$^+$(1) := (\aa$_k$$\oplus$\uu)\www; \hfill}
\ligne{\hskip 10pt \iff{} \aa$_k$ = \uu{} 
\tthen{} list$^-$(1) := \uu\www; list$^+$(1) := \numd\bbb; \endif;\hfill}
\ligne{\hskip 10pt \iff \aa$_k$ = \numd{} 
\tthen{} list$^-$(1) := \numd\bbb; \endif; %\hfill}
\hskip 10pt \iff \aa$_k$ = \ddd{} \tthen{} list$^+$(1) := \zzz\www; \endif;\hfill}
%%%%%%%
\ligne{\hskip 10pt $j$ := 1; first := 1;\hfill}
\ligne{\hskip 10pt \ffor{} $i$ \iin [0..$k$$-$1] {\bf in reverse} \hfill}
\ligne{\hskip 10pt \hskip 10pt \lloop{}\hfill}
\ligne{\hskip 10pt \hskip 10pt \hskip 20pt status := st(list$^-$($j$));\hfill}
%%%%%%%
\ligne{\hskip 10pt \hskip 10pt \hskip 20pt \iff{} \aa$_i$ \iin{} $\{$\numd..\ddd$\}$
\hfill}
\ligne{\hskip 10pt \hskip 10pt \hskip 20pt \hskip 10pt \tthen{} 
\ffor{} $h$ \iin{} [first..$j$] \lloop{} list$^-$(h) := list$^+$(h); \endloop;\hfill}
\ligne{\hskip 10pt \hskip 10pt \hskip 20pt \hskip 10pt \hskip 25pt
first := $j$+1;\hfill}
\ligne{\hskip 10pt \hskip 10pt \hskip 20pt \hskip 10pt \hskip 25pt
list$^-$($j$+1) := \aa$_k$\www;\hfill}
\ligne{\hskip 10pt \hskip 10pt \hskip 20pt \hskip 10pt \hskip 25pt
%%%%
\iff{} ((status = \www\zzz) \eet{} (\aa$_i$ = \ccc)) \oor{} (\aa$_i$ = \ddd)\hfill} 
\ligne{\hskip 10pt \hskip 10pt \hskip 20pt \hskip 10pt \hskip 25pt
\hskip 10pt \tthen{} list$^+$($j$+1) := \zzz\www;\hfill}
\ligne{\hskip 10pt \hskip 10pt \hskip 20pt \hskip 10pt \hskip 25pt
\hskip 10pt \eelse{} list$^+$($j$+1) := (\aa$_i\oplus$\uu)\www; \hfill}
\ligne{\hskip 10pt \hskip 10pt \hskip 20pt \hskip 10pt \hskip 25pt \endif;\hfill}
%%%%
\ligne{\hskip 10pt \hskip 10pt \hskip 20pt \hskip 10pt \hskip 25pt
\iff{} \aa$_i$ = \numd\hfill}
\ligne{\hskip 10pt \hskip 10pt \hskip 20pt \hskip 10pt \hskip 25pt
\hskip 10pt \tthen{} \iff{} status \iin{} $\{$\www,\bbb$\}$\hfill}
\ligne{\hskip 10pt \hskip 10pt \hskip 20pt \hskip 10pt \hskip 25pt
\hskip 10pt \hskip 20pt \hskip 10pt \tthen{} list$^-$($j$+1) := \numd\bbb;\hfill}
\ligne{\hskip 10pt \hskip 10pt \hskip 20pt \hskip 10pt \hskip 25pt \endif;\hfill}
\ligne{\hskip 10pt \hskip 10pt \hskip 20pt \endif;\hfill}
%%%%%%%
\ligne{\hskip 10pt \hskip 10pt \hskip 20pt \iff{} \aa$_i$ \iin{} $\{$\zzz,\uu$\}$ \hfill}
\ligne{\hskip 10pt \hskip 10pt \hskip 20pt \hskip 10pt \tthen{}
\iff{} ((\aa$_i$ = \zzz) \eet{} (status \iin{} $\{$\www\zzz,\www\uu$\}$))\hfill}
\ligne{\hskip 10pt \hskip 10pt \hskip 20pt \hskip 10pt \hskip 40pt
\oor{} ((\aa$_i$ = \uu) \eet{} (status \nnot{} \iin $\{$\www,\uu$\}$))\hfill}
\ligne{\hskip 10pt \hskip 10pt \hskip 20pt \hskip 10pt \hskip 20pt
\hskip 10pt \tthen{}
\ffor{} $h$ \iin{} [first..$j$] \lloop{} list$^+$(h) := list$^-$(h); \endloop;\hfill}
\ligne{\hskip 10pt \hskip 10pt \hskip 20pt \hskip 10pt \hskip 25pt
\hskip 10pt \hskip 20pt
first := $j$+1;\hfill}
\ligne{\hskip 10pt \hskip 10pt \hskip 20pt \hskip 10pt \hskip 25pt
\hskip 10pt \hskip 20pt \iff{} \aa$_i$ = \uu{} \tthen{} list$^-$($j$+1) := \uu\bbb;
\hfill} 
\ligne{\hskip 10pt \hskip 10pt \hskip 20pt \hskip 10pt \hskip 25pt
\hskip 10pt \hskip 20pt \hskip 37pt \eelse{} list$^-$($j$+1) := \zzz\www; \endif;\hfill}
\ligne{\hskip 10pt \hskip 10pt \hskip 20pt \hskip 10pt \hskip 25pt
\hskip 10pt \hskip 20pt list$^+$($j$+1) := \aa$_i\oplus$\uu\www;
\hfill}
\ligne{\hskip 10pt \hskip 10pt \hskip 20pt \hskip 10pt \hskip 20pt
\hskip 10pt \eelse{}  \iff{} \aa$_i$ = \zzz\hfill}
\ligne{\hskip 10pt \hskip 10pt \hskip 20pt \hskip 10pt \hskip 25pt
\hskip 10pt \hskip 20pt \hskip 10pt \tthen{} list$^-$($j$+1) := \zzz\www; 
list$^+$($j$+1) := \uu\bbb;\hfill}
\ligne{\hskip 10pt \hskip 10pt \hskip 20pt \hskip 10pt \hskip 25pt
\hskip 10pt \hskip 20pt \hskip 10pt \eelse{} list$^-$($j$+1) := \uu\bbb;
list$^+$($j$+1) := \numd\www;
\endif;\hfill}
\ligne{\hskip 10pt \hskip 10pt \hskip 20pt \hskip 10pt \hskip 25pt \endif;\hfill}
\ligne{\hskip 10pt \hskip 10pt \hskip 20pt \endif;\hfill}
%%%%%%%
\ligne{\hskip 10pt \hskip 10pt \hskip 20pt $j$ := $j$+1;\hfill}
\ligne{\hskip 10pt \endloop;\hfill}
\vskip 5pt
\hrule height 0.3pt depth 0.3pt width \hsize
}
\hfill}
}
\vskip 10pt
\noindent
and list$^+$ step by step
as follows. Let $\mu$ be the node identified by list$^-(j)$ and $\mu$+1 be the one 
identified by list$^+(j)$. When $j=1$, $\mu$ is identified as the son of the root whose
signature is the highest digit of~{\bf [$\nu$]}. We read the next digit \aa{} 
of~{\bf [$\nu$]}. The 
status of~$\mu$ and~\aa{} allow us to identify the node~$\omega$
such that \aa{} occurs at the right place in the metallic code of nodes which
belong to the tree ${\cal T}_{\omega}$. More precisely, the concerned nodes
lay to the right of the \zzz-branch issued from~${\cal T}_{\omega}$
and to the left of the \zzz-branch issued from~${\cal T}_{\omega+1}$, that latter
branch being included. We can decide which will be the next pair of nodes~$\omega$
and~$\omega$+1 to store in our tables: if \aa{} is not~\uu, we know whether $\omega$
is in ${\cal T}_{\mu}$ or in ${\cal T}_{\mu+1}$. This depends on \aa{} and on the
status of~$\mu$. Algorithm~\ref{apathpp23} carefully scrutinizes the required conditions.
Let us stress the following feature: if $\omega\in{\cal T}_{\mu}$ for instance,
it may happen, this is mostly the case, that $\omega$+1 cannot be reached from
the path leading to~$\mu$+1{} in the tree. In that situation, we decide that 
\hbox{list$^+(h)$ = list$^-(h)$} for \hbox{$h\leq j$}. We may organise the computation
in such a way that we have not to perform that latter identification from the root.
It is enough to remember the last point where such an identification was performed.
This the role of the variable first in the algorithm. To better understand what
may happen, we can note that when \hbox{\aa=\zzz} or \hbox{\aa=\uu}, it is not clear to 
which sub-tree $\nu$ belongs. If $\mu$ is a \www\aa-node, $\nu$ may fall under
the tree rooted at~$\mu$ or in the one rooted at~$\mu$+1. In some cases, that can be
decided in the last digit only: it is the case if \hbox{\bf [$\nu$] $=$ \uu$^{k+1}$\ff}
where \hbox{\ff{} $\in\{\uu,\numd\}$}. Outside such cases, the result of the computation
is to be found in list$^-$, by construction. The interest of this way of computation
is that {\bf [$\nu$]} is read once, without repetition and that each execution of
the body of the \ffor-loop is bounded by a constant, except the updating of
list$^-$ and list$^+$. Also note that in an updating, the new path does not go to
the left of the previous path recorded in list$^-$. Now, thanks to the memorization of 
the last final place of the previous updating, the cumulative effect of the actualization 
process is equivalent to the reading of each table from its lowest index up to its 
highest one. 

As our argument is based on the digits of~{\bf [$\nu$]}, the algorithm may also be
applied to the tiling $\{p,4\}$ without any change. Consequently we proved:

\begin{thm}\label{twpathlin}
	Algorithm~{\rm\ref{apathpp23}} provides an algorithm to compute the
path from the leading tile of a sector to a given tile $\tau$ of the sector in the tiling
$\{p,4\}$ or the tiling $\{p$$+$$2,3\}$ which is linear in time with respect
to the metallic code {\bf [$\nu$]} of the node.
\end{thm}

\subsection{The black metallic tree in the tilings 
$\{\ppp,\numq\}$ and $\{\ppp$\sympl$\numd,\numt\}$}
\label{bpentahepta}
  
    It is time to indicate which place a black metallic tree takes in the tilings
$\{p,4\}$ and $\{p$+$2,3\}$.

    As illustrated by Figure~\ref{fbandes}, the sectors defined by 
Figure~\ref{fspan7_9til} in Sub section~\ref{stilings} can be split
with the help of regions of the tiling generated by the white metallic tree and by the
black one.

In the figure, the sector is split into a tile, we call it the {\bf leading tile}, 
and a complement which can be split into
$p$$-$3 copies of the sector and a region spanned by the black metallic tree which we
call a {\bf strip}.

In both tilings, the strip appears as a region delimited by two lines~$\ell_1$
and~$\ell_2$ which are non-secant. It means that they never meet and that they also are 
not parallel, a property which is specific of the hyperbolic plane. There is a third line 
which supports the side of the tile~$\tau$ which is associated with the root of the 
black metallic tree. That line is the common perpendicular to~$\ell_1$ and~$\ell_2$.
The tile~$\tau$ is called the {\bf leading tile} of the strip.
It is worth noticing that the way we used to split the sector can be recursively repeated
in each sector generated by the process of splitting. We can note that the strip itself
can be exactly split into a tile, a strip and $p$$-$4 sectors, as can easily be seen on 
the 
right-hand side picture of Figure~\ref{fbandes}. This process is closely related
with the generating rules of the metallic trees. 

\vskip 10pt
\vtop{
\ligne{\hfill
\includegraphics[scale=0.6]{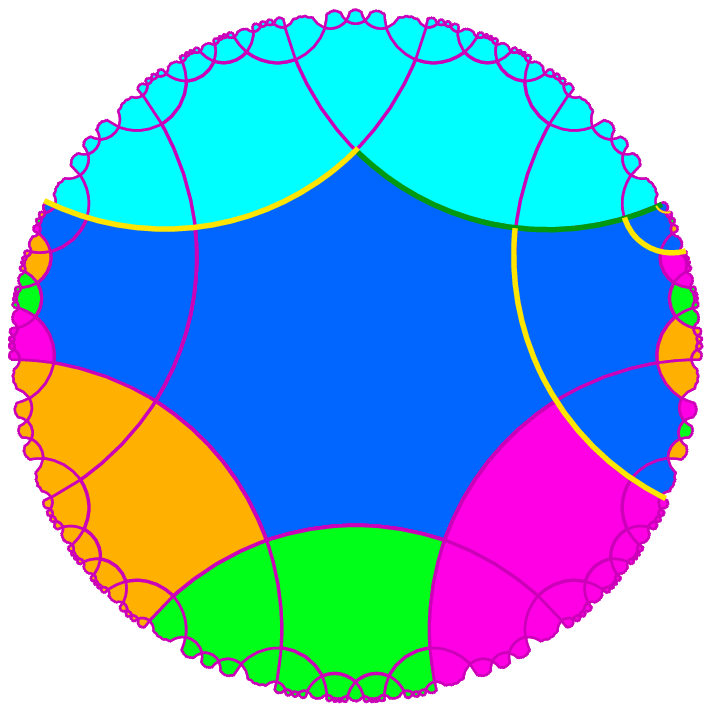}
\includegraphics[scale=0.6]{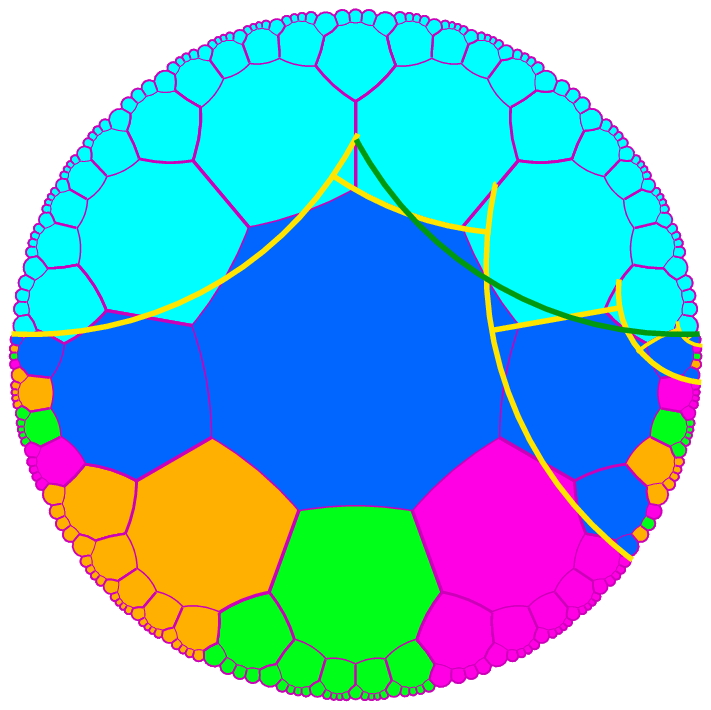}
\hfill}
\vspace{-10pt}
\ligne{\hfill
\vtop{\leftskip 0pt\parindent 0pt\hsize=300pt
\begin{fig}\label{fbandes}
\leurre
The decomposition of a sector spanned by the white metallic tree into a tile, then
two copies of the same sector and a strip spanned by the black metallic tree.
To left: the decomposition in the tiling $\{p,4\}$; to right, the 
decomposition in the tiling $\{p$$+$$2,3\}$.
In both cases, the dark blue colour indicates the black nodes while the white ones
are indicated in dark yellow, in green and in purple.
\end{fig}
}
\hfill}
}
\vskip 5pt
At this point, it can be noticed 
that there are several ways to split a sector and a strip again into strips ans sectors.
This can be associated with other rules for generating a tree which we again call 
a metallic tree. There are still two kinds of nodes, white and black ones. But the
rules are different by the order in which the black son occurs among the sons of a node.
There are \hbox{$p$$-$3} choices for black nodes and \hbox{$p$$-$2} of them for 
white ones. Accordingly,
there are \hbox{$(p$$-$$2)(p$$-$$3)$} possible definitions of a metallic tree. 
We can also decide to
choose which rule is applied each time a node is met. In~\cite{mmJUCStools}
those possibilities are investigated in the case when $p=5$. 
We refer the interested reader to that paper.

\ifnum 1=0 {
\vskip 10pt
\vtop{
\ligne{\hfill
\includegraphics[scale=0.59]{rubans_penta.ps}
\includegraphics[scale=0.5]{rubans_hepta.ps}
\hfill}
\vspace{-10pt}
\ligne{\hfill
\vtop{\leftskip 0pt\parindent 0pt\hsize=300pt
\begin{fig}\label{frubans}
\leurre
The decomposition of a sector spanned by the white Fibonacci tree into a sequence of
pairwise adjacent strips spanned by the black Fibonacci tree.
To left: the decomposition in the pentagrid; to right, the decomposition in the heptagrid.
In both cases, the lines which delimit the strips spanned by the tree.
\end{fig}
}
\hfill}
}
\vskip 5pt
\vskip 10pt
\vtop{
\ligne{\hfill
\includegraphics[scale=0.59]{harpe_penta.ps}
\includegraphics[scale=0.5]{harpe_hepta.ps}
\hfill}
\vspace{-10pt}
\ligne{\hfill
\vtop{\leftskip 0pt\parindent 0pt\hsize=300pt
\begin{fig}\label{fharpes}
\leurre
Another look on the decomposition of the sector given by Figure~{\rm\ref{frubans}}.
The structure of the Fibonacci is erased in order to highlight the decomposition
into pairwise adjacent strips.
\end{fig}
}
\hfill}
}
\vskip 5pt
} \fi

   But a sector can be split in another way which is illustrated by 
figure~\ref{fbandes}. Consider a sector~${\cal S}_0$. Consider its leading tile~$T$.
That tile is associated with the root of the white metallic tree.
Assume that we associate it with the black metallic tree in such a way that
in the association the leftmost son of~$T$ is again the black son of the root in
both trees. What remains in the sector? It remains a node which we can associate
with the root of the white metallic tree. A simple counting argument, taking into
account that the levels are different by one step from the white tree to the black
one in that construction, shows us that in this way we define an exact splitting
of the sector. And so, there is another way to split
the sector: into a strip ${\cal B}_0$ and a sector again, ${\cal S}_1$. Now, what was 
performed 
for~$\cal S$ can be repeated for~${\cal S}_1$ which generates a strip ${\cal B}_1$
and a new sector ${\cal S}_2$. Accordingly, arguing by induction we proved:

\begin{thm}\label{tstrip7_9}
The sector associated to the white metallic tree can be split into a sequence
of pairwise adjacent strips ${\cal B}_n$, \hbox{$n\in\mathbb N$}, associated to the 
black metallic tree. 
Equivalently, the white metallic tree can be split into the union of a sequence of copies
of the black metallic tree. The leading tiles of the ${\cal B}_n$'s are associated
with the nodes \hbox{$M_{n}$} of the white metallic tree, {\it i.e.} the
nodes which are on the rightmost branch of the white metallic tree.
\end{thm}

Note that the proof is a bit easier if it is performed starting from the tree.
The decomposition stated in the theorem is straightforward from the structure
of the rules. Let us remind ourselves that the rule for the white nodes
is \hbox{\bf w $\rightarrow$ bw$^{p-3}$} and that it is
\hbox{\bf b $\rightarrow$ bw$^{p-4}$} for a black node. The difference on the sons
is that the $p$$-$3 first ones are the same for all nodes. A white node appends an
additional node to the rightmost one, so that a white metallic sub-tree of height~$h$+1
can be decomposed into a black metallic sub-tree of height~$h$+1 and a white metallic
sub-tree of height~$h$ whose root is considered as a son of the root of the
black tree of height~$h$+1. Clearly, for infinite trees as considered here, this
splitting gives rise to the theorem.

\subsection{The neighbours of a node of the black metallic tree 
in $\{p,4\}$ and $\{p$+$2,3\}$ under the natural numbering of that tree}
\label{smbneighpathp}

   Let us consider a tile of a sector which falls in the part of it which is spanned by
$\cal B$, the black metallic tree. As the metallic code defined with respect to the
numbering of that tree is different from the one considered in the relations~$(18)$
and~$(19)$, we have to compute appropriate relations.

\begin{lemm}\label{lmbneigh}
Let $\tau$ be a node in the black metallic tree and let 
\hbox{\bf [$\tau$] $=$ a$_k$$..$a$_0$} be its black metallic code.
Let \hbox{\bf [$\tau$]$\ominus$\uu{} $=$ b$_k$$..$b$_0$}. Then, the black metallic
codes of the neighbours of~$\tau$ are given by Table~{\rm\ref{tmbneigh}}.
In the left, right hand-side part of the table, the codes for the neighbours
in the tiling $\{p,4\}$, $\{p$$+$$2,3\}$ respectively.
\end{lemm}

\noindent
Proof. The proof proceeds from the previous study of the sons signature in a black
metallic tree. We take into account that the numbering of the sides is the same as in
Subsections~\ref{smwneighpathp4} and~\ref{smwneighpathpp23}. We have to establish the
correspondence between the number of a side and the signature of the corresponding 
neighbour. In the tiling $\{p,4\}$, the sons of a white node~$\tau$ are its 
$i$-neighbours, with \hbox{$i\in\{2..p$$-$$1\}$}. As the leftmost son of $\tau$
is $(\tau)_2$, the signature of the $i$-neighbour is \hbox{\bf [$i$$-$$2$]},
as the signature of the leftmost son is \zzz. As the signature of the 
leftmost son of a black node is \zzz{} too and as it is 3-neighbour of the node,
we have that \hbox{$i\in\{3,p$$-$$1\}$} and the signature of the $i$-neighbour
is \hbox{\bf [$i$$-$$3$]}. Table~\ref{tmbneigh} gives the whole computations,
for the tiling $\{p,4\}$ in its left hand-side part, for the tiling $\{p$+$2,3\}$
in its right hand-side one. In the table, a $\star$ indicates that the corresponding
neighbour(s) belong(s) to another strip. 

\vtop{
\begin{tab}\label{tmbneigh}
Black metallic codes for the neighbours of a node belonging to the black metallic
tree.
\end{tab}
\vskip -8pt
\ligne{\hfill\vtop{\leftskip 0pt\hsize=250pt
\hrule height 0.3pt depth 0.3pt width \hsize
\vskip 10pt
\ligne{\hskip 50pt in $\{p,4\}$\hfill in $\{p$$+$$2,3\}$\hskip 100pt}
\vskip 5pt
\ligne{\hfill
%  in {p,4}:
$\vcenter{\vtop{\leftskip 0pt\hsize=100pt\bf
\ligne{\hfill \rm if $\tau$ is a \wwaa-node:\hfill}
\vskip 2pt
	\ligne{[$\tau_1$] $=$ b$_k$$..$b$_1$$,$ \hfill}
	\ligne{[$\tau_i$] $=$ b$_k$$..$b$_0$[$i$$-$$2$]$,$\hfill}
        \ligne{\hskip 10pt $\ i\in[2..p$$-$$1],$\hfill}
	\ligne{[$\tau_p$] $=$ a$_k$$..$a$_0$\zzz$,$\hfill}
\vskip 2pt
\ligne{\hfill \rm if $\tau$ is a \www\zzz-node:\hfill}
\vskip 2pt
	\ligne{[$\tau_1$] $=$ b$_k$$..$b$_1$$,$ \hfill}
	\ligne{[$\tau_i$] $=$ b$_k$$..$b$_0$[$i$$-$$2$]$,$\hfill}
        \ligne{\hskip 10pt $\ i\in[2..p$$-$$2],$\hfill}
	\ligne{[$\tau_p$] $=$ [$\tau_1$$+$$1$]$,\ast$\hfill}
\vskip 2pt
	\ligne{\hfill \rm if $\tau$ is a \bbb\zzz-node:\hfill}
\vskip 2pt
	\ligne{[$\tau_1$] $=$ b$_k$$..$b$_1$$,$\hfill}
        \ligne{[$\tau_2$] $=$ [$\tau_1$] $\ominus$ \uu$,$\hfill}
	\ligne{[$\tau_i$] $=$ b$_k$$..$b$_0$[$i$$-$$3$]$,$\hfill}
        \ligne{\hskip 10pt $\ i\in[3..p$$-$$1],$\hfill}
	\ligne{[$\tau_p$] $=$ a$_k$$..$a$_0$\zzz,\hfill}
\vskip 2pt
	\ligne{\hfill \rm if $\tau$ is a \bbb\uu-node:\hfill}
\vskip 2pt
	\ligne{[$\tau_1$] $=$ b$_k$$..$b$_1$$,$\hfill}
        \ligne{[$\tau_2$] $=$ [$\tau_4$$-$$1$]$,\ast$\hfill}
	\ligne{[$\tau_i$] $=$ b$_k$$..$b$_0$[$i$$-$$2$]$,$\hfill}
        \ligne{\hskip 10pt $\ i\in[3..p$$-$$1],$\hfill}
	\ligne{[$\tau_p$] $=$ a$_k$$..$a$_0$\uu,\hfill}
}}
$
\hfill
%  in {p+2,3}:
$\vcenter{\vtop{\leftskip 0pt\hsize=100pt\bf
\ligne{\hfill \rm if $\tau$ is a \wwaa-node:\hfill}
\vskip 2pt
	\ligne{[$\tau_1$] $=$ b$_k$$..$b$_1$$,$ \hfill}
	\ligne{[$\tau_2$] = [$\tau$] $\ominus$ \uu$,$\hfill}
	\ligne{[$\tau_i$] $=$ b$_k$$..$b$_0$[$i$$-$$3$]$,$\hfill}
	\ligne{\hskip 10pt $i\in[3..p],$\hfill}
	\ligne{[$\tau_{p+1}$] $=$ a$_k$$..$a$_0$\zzz$,$\hfill}
	\ligne{[$\tau_{p+2}$] $=$ [$\tau$] $\oplus$ \uu$.$\hfill}
\vskip 2pt
\ligne{\hfill \rm if $\tau$ is a \www\zzz-node:\hfill}
\vskip 2pt
	\ligne{[$\tau_1$] $=$ b$_k$$..$b$_1$$,$ \hfill}
	\ligne{[$\tau_2$] = [$\tau$] $\ominus$ \uu$,$\hfill}
	\ligne{[$\tau_i$] $=$ b$_k$$..$b$_0$[$i$$-$$3$]$,$\hfill}
	\ligne{\hskip 10pt $i\in[3..p$$-$$1],$\hfill}
	\ligne{[$\tau_p$] $=$ a$_k$$..$a$_0$\zzz$,$\hfill}
	\ligne{[$\tau_{p+1}$] $=$ [$\tau_1$$+$$1$]$,\ast$\hfill}
	\ligne{[$\tau_{p+2}$] $=$ $($[$\tau_1$$+$$1$]$)_4,\ast$\hfill}
\vskip 2pt
	\ligne{\hfill \rm if $\tau$ is a \bbb\zzz-node:\hfill}
\vskip 2pt
	\ligne{[$\tau_1$] $=$ b$_k$$..$b$_1$$,$\hfill}
        \ligne{[$\tau_2$] $=$ [$\tau_1$] $\ominus$ \uu$,$\hfill}
	\ligne{[$\tau_3$] $=$ [$\tau$] $\ominus$ \uu$,$\hfill}
	\ligne{[$\tau_i$] $=$ b$_k$$..$b$_0$[$i$$-$$4$]$,$\hfill}
	\ligne{\hskip 10pt $\ i\in[4..p],$\hfill}
	\ligne{[$\tau_{p+1}$] $=$ a$_k$$..$a$_0$\uu$,$\hfill}
	\ligne{[$\tau_{p+2}$] $=$ [$\tau$] $\oplus$ \uu$.$\hfill}
\vskip 2pt
\ligne{\hfill \rm if $\tau$ is a \bbb\uu-node:\hfill}
\vskip 2pt
	\ligne{[$\tau_1$] $=$ a$_k$$..$a$_1$$,$\hfill}
        \ligne{[$\tau_2$] $=$ [$\tau_4$$-$$1$]$,\ast$\hfill}
	\ligne{[$\tau_3$] $=$ [$(\tau_4$$-$$1)_p$]$,\ast$\hfill}
	\ligne{[$\tau_i$] $=$ b$_k$$..$b$_0$[$i$$-$$3$]$,i$\hfill}
        \ligne{\hskip 10pt $i\in[4..p$$-$$2$]$,$\hfill}
	\ligne{[$\tau_{p+1}$] $=$ a$_k$$..$a$_0$\zzz$.$\hfill}
        \ligne{[$\tau_{p+2}$] $=$ [$\tau$] $\oplus$ \uu$,$\hfill}
}}
$
\hfill}
\vskip 10pt
\hrule height 0.3pt depth 0.3pt width \hsize
}
\hfill}
}
\vskip 10pt
That latter point requires some attention. Let $\nu$ be a \www\zzz-node belonging to
${\cal B}_n$ as defined in Theorem~\ref{tstrip7_9}, assuming that we are in the tiling 
$\{p,4\}$. In the white metallic tree containing all trees ${\cal B}_n$, 
$\nu_p\in{\cal B}_{n+1}$. It is not difficult to see, from the construction of the
${\cal B}_n$'s, that the level~$k$ in ${\cal B}_{n+1}$ is the level~$k$ in
${\cal B}_n$. From that observation, we can see that if $\tau$ is a \www\zzz-node
in ${\cal B}_n$, \hbox{$\tau_p=(\tau_1)^{\circ}_4$} where $\tau_1^{\circ}$ is the same 
number as $\tau_1$ for a node in ${\cal B}_{n+1}$. Note that $\tau_p$ is a black node.
Now, if $\sigma=\tau_1^{\circ}$,
we have that $\tau=(\sigma_4)^{\dag}_2$, where $\sigma^{\dag}$ indicates a node in
${\cal B}_n$ which has the same number as the node~$\sigma$ of ${\cal B}_{n+1}$. Such
connections can be checked on Figures~\ref{fmetalblanc} and~\ref{fmetalcomp}.

\vtop{
\begin{algo}\label{ambpath}
Computation of the path from the leading tile of a strip to the tile belonging to
	the strip. The function {\rm st} gives the status of a node.
\end{algo}
\vskip-8pt
\ligne{\hfill
\vtop{\leftskip 0pt\hsize=330pt
\hrule height 0.3pt depth 0.3pt width \hsize
\vskip 5pt
\ligne{\hskip 10pt node := {\bf a$_k$$..$a$_0$};\hfill}
\ligne{\hskip 10pt list$^-$(1) := \aa$_k$\www; list$^+$(1) := \aa$_k$\www; \hfill}
\ligne{\hskip 10pt \iff{} \aa$_k$ = \uu{} 
\tthen{} list$^-$(1) := \zzz\www; list$^+$(1) := \numd\bbb; \endif;\hfill}
\ligne{\hskip 10pt \iff \aa$_k$ = \ddd{} 
\tthen{} list$^-$(1) := \zzz\www; list$^+$(1) := \zzz\www; \endif; \hfill}
\ligne{\hskip 10pt \iff \aa$_k$ \nnot{} \iin{} $\{$\uu,\ddd$\}$ 
\tthen{} list$^-$(1) := (\aa$\oplus$\uu)\www; list$^+$(1) := list$^-$(1); \endif;\hfill}
%%%%%%%
\ligne{\hskip 10pt $j$ := 1; first := $j$+1;\hfill}
\ligne{\hskip 10pt \ffor{} $i$ \iin [0..$k$$-$1] {\bf in reverse} \hfill}
\ligne{\hskip 10pt \hskip 10pt \lloop{}\hfill}
\ligne{\hskip 10pt \hskip 10pt \hskip 20pt status := st(list$^-$($j$));\hfill}
\ligne{\hskip 10pt \hskip 10pt \hskip 20pt \iff{} (status $\not=$ \www\zzz)
\oor{} \eelse{} (\aa$_i$ $\not=$ \zzz)\hfill}
\ligne{\hskip 10pt \hskip 10pt \hskip 20pt \hskip 10pt
\tthen{} \iff{} \aa$_i$ = \ddd\hfill}
\ligne{\hskip 10pt \hskip 20pt \hskip 10pt \hskip 10pt \hskip 30pt 
\tthen{} list$^-$($j$+1) := \zzz\www; list$^+$($j$+1) := list$^-$($j$+1);\hfill} 
\ligne{\hskip 10pt \hskip 20pt \hskip 10pt \hskip 10pt \hskip 30pt 
\eelse{} list$^-$($j$+1) := (\aa$_i\oplus$\uu\www); 
list$^+$($j$+1) := list$^-$($j$+1);\hfill} 
\ligne{\hskip 10pt \hskip 10pt \hskip 20pt \hskip 10pt \hskip 22.5pt \endif;\hfill}
\ligne{\hskip 10pt \hskip 10pt \hskip 20pt \hskip 10pt \hskip 22.5pt
\iff{} first $<$ $j$+1\hfill}
\ligne{\hskip 10pt \hskip 10pt \hskip 20pt \hskip 10pt \hskip 30pt
\tthen{} \ffor{} $h$ \iin{} [first..$j$]\hfill}
\ligne{\hskip 10pt \hskip 10pt \hskip 20pt \hskip 10pt \hskip 30pt \hskip 22.5pt
\hskip 10pt \lloop{} list$^-(h)$ := list$^+(h)$; \endloop;\hfill} 
\ligne{\hskip 10pt \hskip 10pt \hskip 20pt \hskip 10pt \hskip 30pt
\eelse{} first := $j$+2;\hfill}
\ligne{\hskip 10pt \hskip 10pt \hskip 20pt \hskip 10pt \hskip 22.5pt
\endif;\hfill}
\ligne{\hskip 10pt \hskip 10pt \hskip 20pt \hskip 10pt
\eelse{} -$\,$- status = \www\zzz{} and \aa$_i$ = \zzz \hfill}
\ligne{\hskip 10pt \hskip 10pt \hskip 20pt \hskip 10pt \hskip 22.5pt
list$^-$($j$+1) := \zzz\www; list$^+$($j$+1) := \numd\bbb;\hfill}
\ligne{\hskip 10pt \hskip 10pt \hskip 20pt \hskip 10pt \hskip 22.5pt
\iff{} $i$ = 0\hfill}
\ligne{\hskip 10pt \hskip 10pt \hskip 20pt \hskip 10pt \hskip 22.5pt \hskip 10pt
\tthen{} \iff{} \aa$_i$ $\not=$ \zzz\hfill}
\ligne{\hskip 10pt \hskip 10pt \hskip 20pt \hskip 10pt \hskip 22.5pt \hskip 10pt
\hskip 30pt \tthen{} \ffor{} $h$ \iin{} [first..$j$]\hfill}
\ligne{\hskip 10pt \hskip 10pt \hskip 20pt \hskip 10pt \hskip 22.5pt \hskip 10pt
\hskip 30pt \hskip 22.5pt \hskip 10pt \lloop{} list$^-$($h$) := list$^+$($h$);
\endloop;\hfill}
\ligne{\hskip 10pt \hskip 10pt \hskip 20pt \hskip 10pt \hskip 22.5pt \hskip 10pt
\hskip 22.5pt \endif;\hfill}
\ligne{\hskip 10pt \hskip 10pt \hskip 20pt \hskip 10pt \hskip 22.5pt 
\endif;\hfill}
\ligne{\hskip 10pt \hskip 10pt \hskip 20pt \endif;\hfill}
\ligne{\hskip 10pt \hskip 10pt \hskip 20pt $j$ := $j$+1;\hfill}
\ligne{\hskip 10pt \hskip 10pt \endloop;\hfill} 
%%%%%%%
\vskip 5pt
\hrule height 0.3pt depth 0.3pt width \hsize
}
\hfill}
}
\vskip 10pt
If $\sigma$ is the number of the sector where
$\tau$ lies, the new sector is $\sigma\oplus$1 for the \www\zzz-nodes and it
is $\sigma\ominus$1 for the \bbb\uu-nodes. \hfill $\Box$

Table~\ref{tmbneigh} shows us that the determination of the path in the black metallic 
tree is easier than in the case of the white tree. The reason of that simplification
is that if the digit \aa{} is not \zzz, the next digit exactly determines the sub-tree 
where the given node lies because the next digit is in \hbox{\bf \zzz..\ddd} which
corresponds to the sons of a \www\aa-node. If the node to which we arrive is a black node,
we are sure, from the structure of a black metallic tree, that the digit \ddd{} was
not read.

If the digit arriving at a \www\zzz-node~$\nu$ is \zzz, that digit which occurs in
the metallic code at its position in the metallic code of~$\tau$ also occurs in 
the metallic codes of nodes which belong to the sub-tree issued from $\nu$+1 which is
a \bbb\uu-node. As long as \zzz-digits are read, the indetermination between
those consecutive \www\zzz- and \bbb\uu-nodes happens and it is raised by
the first non \zzz-digit or by the fact that all digits of~{\bf [$\tau$]} were
used. This leads us to Algorithm~\ref{ambpath}. \hfill $\Box$

We can state the following property:

\begin{thm}\label{tbpathlin}
Algorithm~{\rm\ref{ambpath}} provides us with an algorithm to compute the
path from the leading tile of a strip to another tile~$\tau$ of the strip
which is linear in time with respect to~{\bf [$\tau$]}.
\end{thm}

As Algorithm~\ref{ambpath} is much simpler than Algorithm~\ref{apathpp23},
Theorem~\ref{tstrip7_9} offers an alternative way to compute the path from a 
tile~$\tau$ to the
leading tile of the sector which contains~$\tau$: we first compute the path from~$\tau$
to the leading tile~$\rho$ of the strip which contains~$\tau$ and then we take the 
part of the rightmost branch of $\cal W$ which goes from the root of~$\cal W$ to~$\rho$.
By numbering the strips ${\cal B}_n$ given by Theorem~\ref{tstrip7_9}, we obtain
an alternative linear algorithm to compute the path from~$\tau$ to the leading tile
of its sector.

\section{Comparing properties of white metallic trees with those of black ones}
\label{scompnumwb}

The last remark which concludes the previous section invites us to compare the
properties stated by Theorems~\ref{tmpreferred} and~\ref{tmblacksucc}.
The first comparison can be made between the rules~$(14)$ with the rules~$(17)$.
For the convenience of the reader, we repeat them right now:

\vskip 5pt
\ligne{\hfill 
$\vcenter{\vtop{\leftskip 0pt\hsize=300pt\bf
\ligne{\hfill
	\bbb\uu$,$\bbb\numd{} $\rightarrow$ \bbb\numd$($\www\aa$)^{p-5}$\www\zzz$,$
	\www\aa{} $\rightarrow$ \bbb\uu$($\www\aa$)^{p-4}$\www\zzz$,$ 
\hfill}
\ligne{\hfill
	\www\zzz{} $\rightarrow$ \bbb\uu$($\www\aa$)^{p-6}$\www\ccc\www\zzz\www\uu$,$
	\www\uu{} $\rightarrow$ \bbb\numd$($\www\aa$)^{p-6}$\www\ddd\www\zzz\www\uu$,$ 
\hfill}
}}$
\hfill$(14)$\hskip 10pt}
\vskip 5pt
\ligne{\hfill
$\vcenter{\vtop{\leftskip 0pt\hsize=300pt\bf
\ligne{\hfill \bbzz{} $\,\,\,\,\,\rightarrow$ \bbzz$($\wwaa$^{)p-4},$\hskip 20pt
\bbuu{} $\,\,\,\,\,\rightarrow$ \bbuu$($\wwaa$)^{p-4},$\hfill}
\ligne{\hfill \wwaa{} $\,\,\,\rightarrow$ \bbzz$($\wwaa$)^{p-3},$\hskip 20pt
\wwzz{} $\rightarrow$ \bbzz$($\wwaa$)^{p-4}$\wwzz$.$\hfill}
}}$
\hfill $(17)$\hskip 10pt}
\vskip 5pt

In both cases, we have two types for the white nodes but for the black nodes,
we have a single type for the white metallic tree and two ones for the black tree.
The difference comes from the fact that the white metallic tree possesses the preferred
son property while the black metallic tree does not. For the black metallic tree we
gave up the term {\it preferred son}, replacing it by {\it successor}. Of course,
the definition of the successor also applies to the white metallic tree and a way
to rephrase Theorem~\ref{tmpreferred} consists in saying that in a white metallic tree,
the successor of each node is its preferred son, where the preferred son, denote
it~{\bf w}$_p$ is given by the following rules: 
\vskip 5pt
\ligne{\hfill\bf b $\rightarrow$ bw$_\ell^{p-5}$w$_p,$
\hskip 10pt w$_\ell$ $\rightarrow$ bw$_\ell^{p-4}$w$_p,$
\hskip 10pt and w$_r$ $\rightarrow$ bw$_\ell^{p-5}$w$_p$w$_r.$
\hfill $(20)$\hskip 10pt}
\vskip 5pt
\noindent
We can say that the preferred son is always a {\bf w}$_r$-node. In a {\bf b}-node and
in a {\bf w}$_\ell$-one it is necessarily the rightmost son, in a {\bf w}$_r$-one
it is the penultimate son, starting from the leftmost son of the node.
We remind the reader that in~$(5)$, we proved that \hbox{$m_{n+1}=b_{n+1}+m_n$}.
\def\st#1{\hbox{\it st}}
\def\succ#1{\hbox{\it succ}}
In the black metallic tree, we can also rephrase Theorem~\ref{tmblacksucc} as follows,
denoting by \st{}$(\nu)$ the status of~$\nu$ and its successor by \succ{}$(\nu)$:
\vskip 5pt
\ligne{\hfill 
$\vcenter{\vtop{\leftskip 0pt\hsize=275pt
\ligne{\st{}$(\nu)$ = \bbb{} {\bf or} \st{}$(\nu)$ = \www\aa{} 
$\Rightarrow$ 
(\succ{}$(\nu) = s_\ell(\nu$+$1)$) \eet{} %,\hfill} 
%\ligne{\st{}$(\nu)$ = \bbb{} {\bf or} \st{}$(\nu)$ = \www\aa{} 
%$\Rightarrow$ \
(st{}$($\succ{}$(\nu))$ = \bbb\zzz),\hfill}
\ligne{\st{}$(\nu)$ = \www\zzz{} $\Rightarrow$ (\succ{}$(\nu) = s_r(\nu)$) \eet{}
%,\hfill} 
%\ligne{\st{}$(\nu)$ = \www\zzz{} $\Rightarrow$ 
(\st{}$($\succ{}$(\nu)$ = \www\zzz).\hfill}
}}$
\hfill $(19)$\hskip 10pt}
\vskip 5pt

%\end{document}
Let us look closer at the difference of the rules by one additional white node in the
rightmost position. Remember that $\cal B$ is identified to the black metallic tree
and that it is dotted with its natural numbering. But the nodes of~$\cal B$ may 
receive another numbering: the number they receive in a white metallic tree as $\cal W$ 
is such a tree. For any node~$\nu\in\cal B$, denote by $\nu_{\cal W}$, $\nu_{\cal B}$ the 
numbers received by $\nu$ in $\cal W$, $\cal B$, respectively in their respective natural
numbering. We can see that
both numberings coincide for the root and for all nodes of level~1, the rightmost
one, $\sigma$, excepted which is the root of~$\cal C$, not in $\cal B$,
see Figure~\ref{fmetalcomp}.
Let $\varphi_k$ be the rightmost node of $\cal B$ on the level~$k$ and let
$\lambda_k$ be the leftmost node on the same level. It is not difficult to
see that for any node $\nu$ of $\cal B$ and on level~2, 
\hbox{$\nu_{\cal W}=\nu_{\cal B}$+1}. Indeed, we have 
\hbox{$(\lambda_2)_{\cal B}=(\varphi_1)_{\cal B}$+1} while
\hbox{$(\lambda_2)_{\cal W}=M_1$+1} as proved in Theorem~\ref{theadlevelwb}.
Accordingly, on level~3, for any node~$\nu$ of~$\cal B$, we have
\hbox{$\nu_{\cal W}=\nu_{\cal B}$+$M_1$}. Say that the numbers in $\cal W$ are
shifted by~$M_1$ with respect to those in $\cal B$. We noticed that on level~2 the 
shift was~1,
so that the shift increased by~$m_1$ from level~2 to level~3 and $m_1$
is the number of nodes of \hbox{${\cal W}\backslash{\cal B}$} on level~2,
{\it i.e.} the nodes of~$\cal C$ on its level~1.

\vskip 10pt
\vtop{
\ligne{\hfill
\includegraphics[scale=0.35]{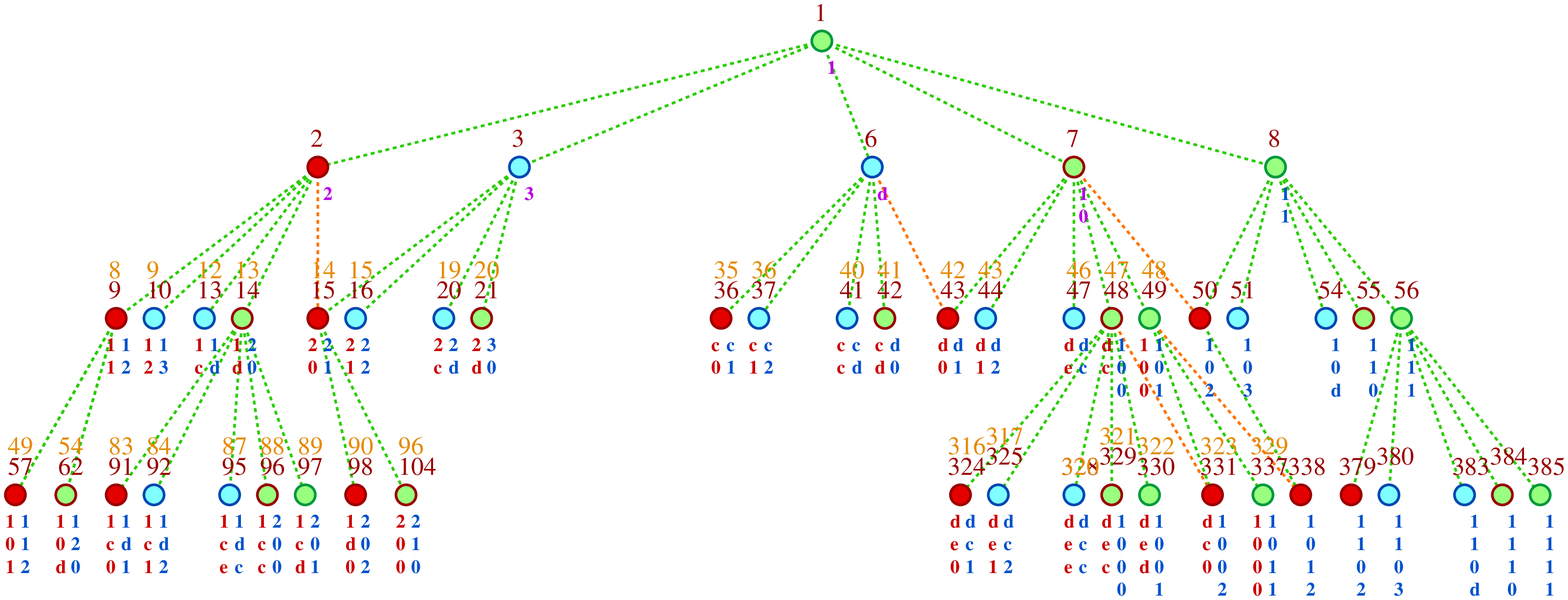}
\hfill}
\vspace{-10pt}
\ligne{\hfill
\vtop{\leftskip 0pt\parindent 0pt\hsize=300pt
\begin{fig}\label{fmetalcomp}
\leurre
Comparing the white metallic numbering, in blue in the figure, on the black metallic 
tree with the natural numbering of that latter one, in red. 
Partial representation of the first three levels of the tree when $p=9$
with the conventions mentioned for Figures~{\rm \ref{fmetalblanc}}
and~{\rm \ref{fmetalnoir}}.
\end{fig}
}
\hfill}
}
\vskip 5pt
Accordingly, by induction on~$n$, assume that on the level~$n$, for any node~$\nu$ 
of~$\cal B$ we have \hbox{$\nu_{\cal W}=\nu_{\cal B}$+$M_{n-2}$}.
Accordingly, \hbox{$(\varphi_n)_{\cal W}=(\varphi_n)_{\cal B}$+$M_{n-2}$},
so that we have 
\hbox{$(\lambda_{n+1})_{\cal W}=(\lambda_{n+1})_{\cal B}$+$M_{n-2}$+$m_{n-1}$},
as the number of nodes of \hbox{${\cal W}\backslash{\cal B}$} on the level~$n$
is $m_{n-1}$, the number of nodes of $\cal C$ on its level~$n$$-$1. Now, from
the equality
\hbox{$M_{n-2}$+$m_{n-1}=M_{n-1}$} which proves our claim, we can state:

\begin{lemm}\label{lcompnumwb}
For any node $\nu$ on the level~$n$$+$$1$ of~$\cal B$, we have:
\vskip 5pt
\ligne{\hfill $\nu_{\cal W}=\nu_{\cal B}$$+$$M_{n-1}$\hfill $(19)$\hskip 10pt}
\end{lemm}

\def\zB{\hbox{$\beta_{\zzz,\cal W}$}}
As an application, let us look at the \zzz-branch of~$\cal W$ and denote it
\zB. The nodes of the 
rightmost branch of $\cal B$ have the metallic code \uu\zzz\uu$^n$ in $\cal W$. 
Accordingly, \zB{} is contained in $\cal B$. 
Accordingly, if $\pi_n$ is the node of~\zB{} which lies on the level~$n$ of $\cal W$,
we have from~$(19)$ that
\hbox{${\pi_n}_{\cal B} = {\pi_n}_{\cal W}-M_{n-2}$}. 
Performing \hbox{\bf \uu\zzz$^n$ $-$ \uu$^{n-2}$} thanks to 
Algorithms~\ref{acomparemc}, \ref{acomplement},
\ref{aaddmetal} and \ref{alibre} we get: 

\begin{cor}\label{lzzzbinB}
Let $\pi_n$ denote the sequence of nodes defined by the following
conditions: $\pi_0$ is the root of $\cal W$, $\pi_{n+1}$ is the penultimate
son of $\pi_n$ for all positive integer~$n$. We have that
\hbox{\bf [${\pi_n}_{\cal W}$] $=$ \uu\zzz$^n$} and
\hbox{\bf [${\pi_n}_{\cal B}$] $=$ \ddd$($\ccc$^-)^{n-2}$\ccc}.
\end{cor}

We already proved that in the black metallic tree, the rightmost node on the level~$n$+1
is numbered~$m_{n+1}$. It was proved by removing from~$M_{n+1}$ the number of nodes
in $\cal C$.
The number of the removed nodes is $M_n$ which repeats the argument used in
the proof of Theorem~\ref{tmpreferred} to show that $\pi_{n+1}$ is the
penultimate son of~$\pi_n$. This confirms the fact that \zB{} is not the rightmost
branch of $\cal B$ as already mentioned.
Note that the sub-tree of~$\cal B$ whose root is
the rightmost son of~$\cal B$ on its level~2 is isomorphic to~$\cal C$. In terms
of the sub-tilings generated by the trees in the tiling $\{p,4\}$ or
$\{p$+$2,3)\}$ that isomorphism corresponds to a shift. In $\cal W$ 
the leftmost black node puts the successor at its expected place in both trees so that 
the white nodes, which have $p$$-$2 sons keep the place of the successors where they 
should be. As the shift which allows us to pass from one numbering to the other
corresponds to a number of nodes in the white tree at the previous level, 
the occurrence of the pattern \zzz$^k$ replacing the forbidden \ddd\ccc$^{k-2}$\ddd{} 
occurs at a right place with respect to the nodes of the previous level.

And so, we have the explanation of the
differences we noticed on the rules $(14)$ and~$(17)$.

\section*{Conclusion}\label{conclude}

    We can conclude the paper with several remarks.

    The first one is that the results of the paper are different from those
of~\cite{mmgsJUCS} as in that paper, the rules place the black son at a very different 
place from the place defined by~$(20)$. Now, the existence of a linear algorithm
constructing the path from a node to the root of the tree was proved in~\cite{mmbook2} 
which takes the setting of~\cite{mmgsJUCS}. Accordingly, the result of the
present paper confirms the result of~\cite{mmbook2}. Now, neither in~\cite{mmgsJUCS}
nor in~\cite{mmbook1}, nor in~\cite{mmbook2} the case of the black tree was investigated
for Fibonacci trees. There is a mention in~\cite{ImaIwaMorita} of the value of $b_n$
in the case of the Fiboancci tree but no other property of its natural numbering, 
in particular no connection with the code associate to that latter numbering. Such 
properties were first studied by the author in~\cite{mmblacktree}
as mentioned in the Introduction.

   The interest of the black metallic tree is confirmed by the general case investigated
in the present paper. We could conclude in~\cite{mmblacktree} that the white Fibonacci
tree is the best tree for navigation purpose in the pentagrid
and in the heptagrid, those tessellation that live in the hyperbolic plane. 
Probably, it is still the case in the general setting considered in the present paper.
However, the simplicity of Algorithm~\ref{ambpath} a bit tempers the previous statement.
For investigations in the black metallic tree only, considering the natural numbering 
of that tree would perhaps be the best issue: there is no difference between black and
white nodes for the determination of the successor of a node, except for the rightmost 
branch of the tree. Now, that rightmost branch is exceptional too in the white metallic
tree. Worse, that exceptional situation is translated to the sub-trees of the white tree.
It is not the case in the black metallic tree equipped with its natural numbering.
Moreover, $\cal B$ covers a large part of~$\cal W$. Also note that a branch passing 
through a given node is unique in a tree. Accordingly, the path joining a node placed 
in a ${\cal B}_n$ tree with \hbox{$n>0$} is the same path as in $\cal W$
to get connected with the root. In particular, the path defined in $\cal W$ passes 
through the same nodes of the rightmost branch of~$\cal W$ as above mentioned. Those
considerations reinforce the interest of Algorithm~\ref{ambpath}.
And so, in this context, the nice property of the 
preferred son which holds in the
white metallic tree and which no more holds for the black one, is replaced by 
a nice property too for the successor of a node.

    Up to a point, the present paper closes the problem for the tilings $\{p,4\}$
and $\{p$+$2,3\}$ for what is the rules~$(1)$. As already noted in~\cite{mmJUCStools},
other rules giving rise to the same number of branching for black and white nodes
as in~$(1)$ could be investigated, we made hint to that feature in
Sub-section~\ref{bpentahepta}. In that paper, we had 6 pairs of rules.
Some of them could give rise to stronger distortions
with respect to the regularity shown in the present paper. Now, in \cite{mmJUCStools},
another question was considered: on each level of the tree, the rules applied 
to a black, white node is taken at random for the possible rules for that type of node.
Little was indicated in~\cite{mmJUCStools} about that situation. In the situation
investigated in the present paper, we have \hbox{$(p$$-$$3)(p$$-$$2)$} possible pairs of
rules, and so, if the rules are taken at random, the number of possibilities is much higher
than in the situation considered in~\cite{mmJUCStools}.
Accordingly, there are still open problems
in the topic considered in the present paper.


\begin{thebibliography}{5}
\newcount\bibi\bibi=1

\bibitem{ImaIwaMorita}
C. Iwamoto, T. Andou, K. Morita, K. Imai,
Computational Complexity in the Hyperbolic Plane, 
{\it Lecture Notes in Computer Science}, {\bf 2420}, (2002), 
Proceedings of {\bf MFCS'2002}, 365-374.

\bibitem{mmJUCStools}
M. Margenstern,
New Tools for Cellular Automata of the Hyperbolic Plane,
{\it Journal of Universal Computer Science},
{\bf 6}(12), (2000), 1226--1252.

\bibitem{mmbook1}
M. Margenstern,
{\it Cellular Automata in Hyperbolic Spaces, vol. $I$, Theory},
Collection: {\it Advances in Unconventional Computing and Cellular Automata},
Editor: Andrew Adamatzky,
Old City Publishing, Philadelphia, (2007), 424p.

\bibitem{mmbook2}
M. Margenstern,
{\it Cellular Automata in Hyperbolic Spaces, vol. $II$, Implementation and
computations},
Collection: {\it Advances in Unconventional Computing and Cellular Automata},
Editor: Andrew Adamatzky,
Old City Publishing, Philadelphia, (2008), 360p.

\bibitem{mmblacktree}
M. Margenstern,
About Fibonacci trees $-$ I $-$, {\it arXiv}:1904.12135, cs.DM.

\bibitem{mmgsJUCS}
M. Margenstern, G. Skordev,
Fibonacci Type Coding for the Regular Rectangular Tilings of the
Hyperbolic Plane,
{\it Journal of Universal Computer Science},
{\bf 9}(5), (2003), 398-422.

\end{thebibliography}
\end{document}